%% file: 2024-www-fp-faith-cr.tex
\documentclass[sigconf,natbib=true,anonymous=false,review=false]{acmart}

\AtBeginDocument{%
  \providecommand\BibTeX{{%
    \normalfont B\kern-0.5em{\scshape i\kern-0.25em b}\kern-0.8em\TeX}}}

\usepackage{latexsym}
\usepackage{graphicx}
\graphicspath{{./images/}}
\usepackage{booktabs} 
\usepackage{color}  
\usepackage{amsmath}  
\usepackage{subcaption}
\usepackage{caption}
\usepackage{tikz}
\usepackage{colortbl} 
\usepackage{framed}
\usepackage{multirow}
\usepackage{multicol}
\usepackage{hyperref}
\usepackage{url}
\usepackage{balance}
\usepackage{verbatim}
\usepackage{cancel}
\usepackage{xspace} 
\usepackage[ruled,vlined]{algorithm2e}
\usepackage{bbold} 
\usepackage{comment}

\newcommand{\struct}[1]{\texttt{\small #1}}
\newcommand{\utterance}[1]{\textit{#1}}
\newcommand{\phrase}[1]{\textit{``#1''}}

\newenvironment{Snugshade}[1][236,236,236]{
    \setlength{\itemsep}{0pt}
     \setlength{\parsep}{0pt}
     \setlength{\topsep}{0pt}
     \setlength{\partopsep}{0pt}
     \setlength{\leftmargin}{1.5em}
     \setlength{\labelwidth}{0em}
     \setlength{\labelsep}{0em} 
    \setlength{\parskip}{0pt}
    \definecolor{shadecolor}{RGB}{#1}%
    \begin{snugshade}
}{%
    \end{snugshade}%
}

\newcommand{\explaignn}{\textsc{Explaignn}\xspace}

\newcommand{\timequestions}{\textsc{TimeQuestions}\xspace}

\newcommand{\clocq}{\textsc{Clocq}\xspace}

\newcommand{\exaqt}{\textsc{Exaqt}\xspace}
\newcommand{\tempoqr}{\textsc{TempoQR}\xspace}
\newcommand{\cronkgqa}{\textsc{CronKGQA}\xspace}

\newcommand{\unikqa}{\textsc{Unik-Qa}\xspace}
\newcommand{\uniqorn}{\textsc{Uniqorn}\xspace}

\newcommand{\tsf}{{TSF}\xspace}

\newcommand{\temporalqa}{\textsc{Faith}\xspace}
\newcommand{\faith}{\textsc{Faith}\xspace}
\newcommand{\benchmark}{\textsc{Tiq}\xspace}
\newcommand{\tiq}{\textsc{Tiq}\xspace}

\newcommand{\fid}{\textsc{FiD}\xspace}

\copyrightyear{2024}
\acmYear{2024}
\setcopyright{rightsretained}
\acmConference[WWW '24]{Proceedings of the ACM Web Conference 2024}{May 13--17, 2024}{Singapore, Singapore}
\acmBooktitle{Proceedings of the ACM Web Conference 2024 (WWW '24), May 13--17, 2024, Singapore, Singapore}

\usepackage{etoolbox}
\makeatletter
\patchcmd{\maketitle}{\@copyrightpermission}{
\begin{minipage}{0.3\columnwidth}
\href{http://creativecommons.org/licenses/by/4.0/}{\includegraphics[width=0.90\textwidth]{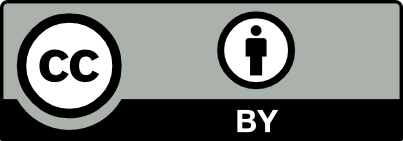}}
\end{minipage}\hfill
\begin{minipage}{0.7\columnwidth}
\href{http://creativecommons.org/licenses/by/4.0/}{This work is licensed under a Creative Commons Attribution International 4.0 License.}
\end{minipage}

\vspace{5pt}
}{}{}

\makeatother

\begin{document}


\title{{Faithful Temporal Question Answering} \mbox{over Heterogeneous Sources}}

\author{Zhen Jia}
\affiliation{%
  \institution{Southwest Jiaotong University}
  \city{Chengdu}
    \country{China}
}
\email{zjia@swjtu.edu.cn}

\author{Philipp Christmann}
\affiliation{%
  \institution{Max Planck Institute for Informatics\\Saarland Informatics Campus}
  \city{Saarbruecken}
 \country{Germany}
}
\email{pchristm@mpi-inf.mpg.de}

\author{Gerhard Weikum}
\affiliation{%
  \institution{Max Planck Institute for Informatics\\Saarland Informatics Campus}
  \city{Saarbruecken}
    \country{Germany}
}
\email{weikum@mpi-inf.mpg.de}

\renewcommand{\shortauthors}{Zhen Jia, Philipp Christmann, \& Gerhard Weikum}

\newcommand{\squishlist}{
    \begin{list}{$\bullet$}{ 
        \setlength{\itemsep}{0pt}
        \setlength{\parsep}{1pt}
        \setlength{\topsep}{1pt}
        \setlength{\partopsep}{0pt}
        \setlength{\leftmargin}{1.5em}
        \setlength{\labelwidth}{1em}
        \setlength{\labelsep}{0.5em} 
    } 
}

\newcommand{\squishend}{
  \end{list}  }
  
\newcommand{\GW}[1]{{\tt\color{blue}{GW: #1}} }
\newcommand{\PC}[1]{{\color{orange}{PC: #1}} }
\newcommand{\todo}[1]{{\color{red}{TODO: #1}} }
\newcommand{\ZJ}[1]{{#1}}
\newcommand{\ZJadded}[1]{{{#1}}}

\newcommand{\myparagraph}[1]{\noindent \textbf{#1}.}
\setcounter{secnumdepth}{4}

\input{sections/00-abstract}

%
%



\maketitle

\input{sections/01-introduction}
\input{sections/02-concepts}
\input{sections/03-method}

\input{sections/04-benchmark}

\input{sections/06-results}

\input{sections/07-related}
\input{sections/08-conclusion}



\bibliographystyle{ACM-Reference-Format}
\bibliography{temporalqa}

\appendix
\input{sections/10-appendix}

\end{document}

%% file: sections/00-abstract.tex
\begin{abstract}

Temporal question answering (QA) 
involves time constraints, with phrases such as \phrase{\dots in 2019} or \phrase{\dots before COVID}. In the former, time is an {\em explicit} condition, in the latter it is {\em implicit}.
%
State-of-the-art methods 
have limitations along three dimensions. 
First, 
with
neural inference,
time constraints
are merely soft-matched,
giving room to invalid or inexplicable answers.
Second, questions with implicit time are poorly supported.
Third, 
answers come from a single source:
either a knowledge base 
(KB) 
or a text corpus.
%
%
%
We propose a temporal QA system that addresses these shortcomings.
First, it 
enforces temporal constraints
for \textit{faithful answering} with tangible evidence.
Second, it 
properly handles {\em implicit questions}.
Third, it operates over \textit{heterogeneous sources},
covering KB, text and web 
tables in a unified manner.
The method has three stages:
(i) understanding the question and its temporal conditions,
(ii) retrieving
evidence from all sources,
and (iii) faithfully answering the question.
As implicit questions are 
sparse in prior benchmarks,
we introduce a principled method for generating diverse questions.
%
Experiments 
show
superior performance over a suite of baselines.
\end{abstract}


%% file: sections/01-introduction.tex
\begin{figure*} [t]
     \includegraphics[width=0.8\textwidth]
     {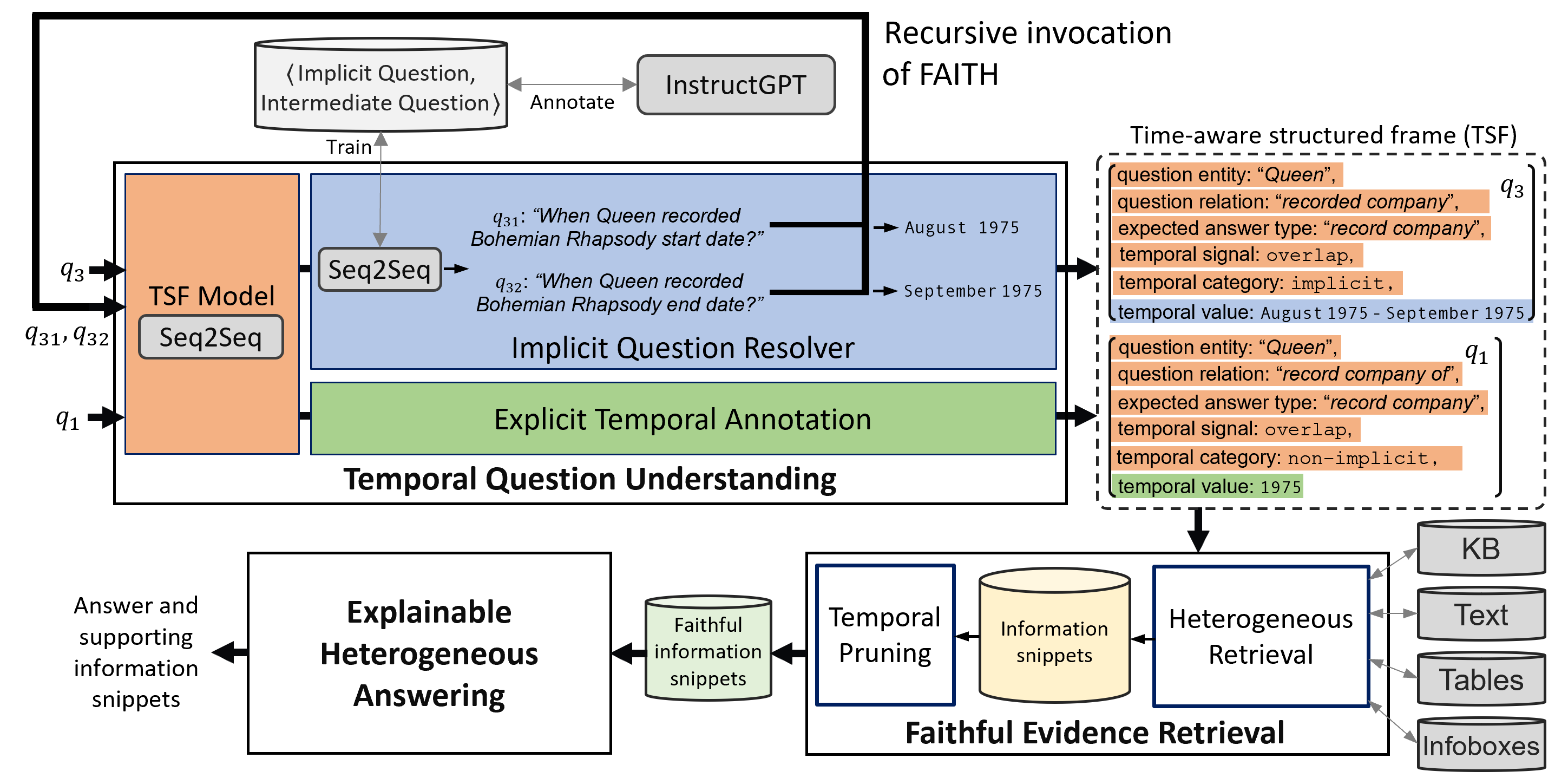}
     \vspace*{-0.3cm}
     \caption{Overview of the \faith pipeline.
     The figure illustrates the process for answering 
     $q_3$
     (\phrase{Queen’s record company when recording Bohemian Rhapsody?})
     and $q_1$
     (\phrase{Record company of Queen in 1975?}).
     For answering $q_3$, two intermediate questions $q_{31}$ and $q_{32}$ are generated,
     and run recursively through the entire temporal QA system.}
     \label{fig:overview}
     \vspace*{-0.1cm}
\end{figure*}

\section{Introduction}

\myparagraph{Motivation}
Question answering (QA) comprises a spectrum of settings for satisfying users' information needs,
ideally giving crisp, entity-level answers to natural-language utterances \cite{saharoy2022question}.
Temporal QA specifically focuses on questions with {temporal conditions}
(e.g., \cite{jia2021complex, saxena2021question, mavromatis2022tempoqr}),
making up a substantial portion of user needs~\cite{zhang2010learning},
but poses challenges that are not properly met by universal QA systems.
Consider the following
example:
\begin{Snugshade}
    $q_1$: \utterance{Record company of Queen in 1975?}
\end{Snugshade}
The band \struct{Queen} had different record companies over the years, so
it is decisive 
to
consider the \textit{explicit temporal constraint} (\phrase{in 1975}).
%
Other questions with explicit time are lookups of dates, such as:
\begin{Snugshade}
    $q_2$: \utterance{When was Bohemian Rhapsody recorded?}
\end{Snugshade}

Another -- underexplored and most challenging -- situation is when questions involve 
\textit{implicit temporal constraints}. These can involve the need to compare different time points or intervals, even when the user input does not explicitly state it. Examples are:
\begin{Snugshade}
    $q_3$: \utterance{Queen's record company when recording Bohemian Rhapsody?}\\
    \indent $q_4$: \utterance{Queen's lead singer after Freddie Mercury?}
\end{Snugshade}
%
For $q_4$, the system has to find out when Mercury died or left the band, in order to compute the correct answer that Brian May (the band's guitarist) took over as lead singer.

The research literature on temporal QA is substantial, including \cite{jia18tequila,jia2021complex,jiao2022improving,saxena2021question,xiao2022modeling,chen2022temporal,ding2022semantic,mavromatis2022tempoqr,chen2021dataset}. Most methods address
all kinds of temporal questions, but are typically less geared for implicit questions.
Some methods operate over curated knowledge bases (KBs) (e.g., \cite{jia18tequila, jia2021complex, ding2022semantic}),
while others are designed for processing text corpora such as news collections or Wikipedia full-text (e.g., \cite{chen2021dataset, ning2020torque}).



\vspace*{0.1cm}
\myparagraph{State-of-the-art limitations}
We observe three major issues:

\squishlist
\item[(i)]
Many methods 
use ``soft-matching'' techniques, based on latent embeddings or language models.
This may lead to invalid answers,
where the non-temporal part of a question is matched, but the temporal constraint is violated.
For example, 
a question about \phrase{Queen's record company in 1990?} may erroneously return \struct{EMI} instead of the correct value \struct{Parlophone}, 
because \struct{EMI} is 
more prominent and was Queen's company on most albums.
Even when the output 
is correct, this could
be by the prominence of the answer alone.
For example, \phrase{Who was Queen's lead singer in 1975?} could return the most popular \struct{Freddie Mercury} without checking the time.
When we vary the question into \phrase{\dots in 2000?},
many systems would still yield \struct{Freddie Mercury}, although he was dead then.
%
This indicates that the system 
has 
incomplete inference and is 
unable to explain its answer derivation. We call this phenomenon {\em unfaithful QA}.
\item[ii)]
A weak spot of temporal QA systems is the handling of {\em implicit questions}.
These are infrequent in established benchmarks.
Some methods~\cite{jia18tequila,ding2022semantic,neelam-etal-2022-sygma} aim to transform the implicit conditions into explicit temporal constraints, based on classifying phrases starting with ``during'', ``before'' etc.
However, they heavily rely on hand-crafted rules 
which are rather limited in scope and cannot robustly handle unforeseen utterances.
\item[(iii)]
Prior methods run on a \textit{single information source}: either a KB {or} a text corpus.
This 
limits QA
coverage: KBs 
are incomplete and 
lack refined detail about events, whereas text collections 
are harder to extract answers from and often 
fail on complex questions
\cite{ding2022semantic, christmann2022beyond}.
%
QA over heterogeneous sources, including also web tables, has been addressed by \cite{oguz2021unikqa,christmann2023explainable},
but these methods 
do not support 
temporal conditions. 
\squishend

\vspace*{0.1cm}
\myparagraph{Approach}
To overcome these limitations, 
we propose \temporalqa
(\underline{FAI}thful \underline{T}emporal question answering over \underline{H}eterogeneous sources),
a temporal QA system that operates over \textit{heterogeneous} sources, seamlessly combining a KB, a text corpus
and web tables. 
Inspired by the architecture of~\cite{christmann2023explainable},
\temporalqa consists of three main stages: 
\squishlist
\item[(i)] \textbf{Temporal Question Understanding} for representing the question intent into a structured frame, with specific consideration of the
temporal aspects;
\item[(ii)] \textbf{Faithful Evidence Retrieval} for identifying relevant pieces of evidence from 
KB, text and tables,
with time-aware filtering to
match the temporal conditions;
\item[(iii)] \textbf{Explainable Heterogeneous Answering} to compute entity-level answers and supporting evidence for explanation. 
\squishend

\vspace*{0.1cm}
A key novelty in the question understanding
is that implicit constraints are resolved into explicit temporal values by generating
intermediate questions and recursively calling \faith itself.
For example, the implicit condition \phrase{when recording Bohemian Rhapsody} in $q_3$
is transformed into  \phrase{when Queen recorded Bohemian Rhapsody?}, and the recursive invocation of \faith returns the explicit condition
\struct{August\,1975\,-\,September\,1975}.
This derived explicit condition is then used in a similar vein as the explicit condition \struct{1975} in $q_1$,
making it easier to answer the information need.
Note that this is not just question rewriting,
but is driven by the
full-fledged
QA system itself over the full suite of heterogeneous sources.

A second key novelty is that, in contrast to most prior works including large language models,
\temporalqa provides \textit{tangible provenance} for the answer derivation.
By providing users with explanatory evidence for answers, \faith is a truly faithful temporal QA system.

Existing benchmarks for temporal QA focus on a single information source at hand (either a KB or a text corpus),
and include only few questions with implicit constraints (so the weak performance on these hardly affects the overall
results).
Therefore, we devise a new method for automatically creating temporal questions with {\em implicit constraints}, with systematic controllability of different aspects, including
the relative importance of different source types (text, infoboxes, KB),
coverage of topical domains (sports, politics etc.),
fractions of prominent vs. long-tail entities,
question complexity, and more.
This way, we construct a new dataset named \benchmark with $10{,}000$
questions and
answers accompanied by supporting evidence.

Our code and data is
available at
\textbf{\url{https://faith.mpi-inf.mpg.de}}.

\vspace*{0.1cm}

\myparagraph{Contributions}
The salient contributions of this work are:
\squishlist
\item
the first temporal QA system that 
taps into
heterogeneous 
sources, and 
gives 
faithful answers with explanatory evidence;
\item
a mechanism that transforms implicit temporal constraints into explicit conditions, by recursively calling the QA system itself;
\item a principled method for automatic construction of diverse and difficult temporal questions, 
releasing the \benchmark
benchmark.
\squishend

%% file: sections/02-concepts.tex
\section{Concepts and Notation}
\label{sec:concepts}
This section introduces salient concepts and notation
for this work.

\myparagraph{Temporal value}
A \textit{temporal value} indicates a point in time or time interval. 
It
can be a
specific \textit{date} (e.g., \struct{24\,November\,1991}),
a \textit{year} (e.g., \struct{1975}),
or a {time period} (e.g., \struct{August\,1975\,-\,September\,1975}).


\myparagraph{Temporal constraint}
A \textit{temporal constraint} specifies a 
condition about a time point or interval that
has to be satisfied by the answer and its evidence.
Temporal constraints consist of a temporal value, and a \textit{temporal signal}
(like\,\textit{before}, \textit{after}, \textit{overlap}).
An example of a (verbalized) temporal constraint is \phrase{in 1970}.

\myparagraph{Explicit question}
An \textit{explicit question} mentions a specific temporal constraint explicitly,
as in \phrase{Record company of Queen in 1975?}.

\myparagraph{Implicit question}
An \textit{implicit question} also specifies a temporal constraint, but keeps this constraint \textit{implicit}
without mentioning the actual temporal value:
\phrase{Queen’s record company when recording Bohemian Rhapsody?}.

\myparagraph{Answer}
An \textit{answer} to a question is either an entity (e.g., \struct{Brian\,May}) or a literal such as a date (e.g., \struct{24\,August\,1975}), year (e.g., \struct{1975}) or number (e.g., \struct{3}).

\myparagraph{Evidence}
An \textit{evidence} is given 
with an answer as explanatory support.
The evidence consists of \textit{information snippets} that are retrieved from a KB, a text corpus, a table, or a Wikipedia infobox.
Following ~\cite{christmann2022conversational}, we consider 
snippets on a sentence-level:
text is split into sentences, and KB-facts, table rows and infobox entries
are \textit{verbalized} by concatenating the individual pieces.

\myparagraph{Faithfulness}
A system answers a question \textit{faithfully} if its evidence, provided with the answer, contains:
(i) {the answer}, 
(ii) {all entities that appear in the question (with any surface name)},
(iii) {all predicates that appear in the question (at least in paraphrased or implicit form)},
(iv) {a temporal expression that satisfies the temporal constraint of the question}.
The first three aspects are valid in the context of any QA system; the fourth is specific to temporal QA.

%% file: sections/03-method.tex
\section{FAITH Method}
\label{sec:method}
Fig.~\ref{fig:overview} provides an overview of the system architecture, illustrated with the processing of the running examples $q_3$ and $q_1$. The following subsections present the three main components
(understanding, retrieval, and answering), and will refer to these examples.

\subsection{Temporal Question Understanding}
\label{subsec:tqu}
The goal of this first stage is to capture the temporal information need in a frame-like structure.
Notably, this stage  
identifies and categorizes temporal constraints in the user input,
which is later used for pruning temporally-inconsistent answer candidates.

\myparagraph{\tsf}
Inspired by~\cite{christmann2022conversational}
and~\cite{ho2019qsearch} (both addressing other, non-temporal, kinds of QA),
we propose to learn a \textit{\underline{T}ime-aware \underline{S}tructured \underline{F}rame (\tsf)} for an incoming temporal question. The {\tsf} includes both general-QA-relevant slots:
\squishlist
\item \textit{question entity}, 
\item \textit{question relation},
\item \textit{expected answer type},
\squishend
\noindent and temporal-QA-relevant slots:
\squishlist
\item \textit{temporal signal}, indicating the kind of temporal relation,
\item \textit{temporal category}, indicating the type of temporal constraint,
\item \textit{temporal value},
the time point or interval of interest (if present).
\squishend

The \textit{question entity} and \textit{relation} are taken from the surface form of the question
(i.e. \textit{not} linked to KB) to allow for uniform treatment of heterogeneous sources.
The \textit{expected answer type} is learned from the training data, in which 
the KB-type of the gold answer is used. 
The \textit{temporal signal} can be
\struct{overlap} (e.g., from cues like \phrase{in}, \phrase{during}),
\struct{before} (e.g., from cues like \phrase{before}, \phrase{prior to}),
or \struct{after} (e.g., from cues like \phrase{after}, \phrase{follows}).
We categorize the constraint into \textit{implicit} (e.g., $q_3$ and $q_4$) and \textit{non-implicit} (e.g., $q_1$ and $q_2$).
The \textit{temporal value} 
can be a 
\textit{year}, 
\textit{date} or \textit{time period}. 
Both the temporal signal and value are derived by identifying and normalizing key phrases in the input question. For example, the {\tsf} for $q_1$ is:
\begin{Snugshade}
$\langle$
    {question entity}: \phrase{Queen},\\
    \indent {question relation}: \phrase{record company of},\\
    \indent expected answer type: \phrase{record company},\\
    \indent temporal signal:  \struct{overlap},\\
    \indent temporal category: \struct{non-implicit},\\
    \indent temporal value: \struct{1975}
$\rangle$
\end{Snugshade}
\noindent Note that in case the question does not specify temporal constraints (e.g., $q_2$), the respective fields are simply kept empty.

\myparagraph{Resolving implicit questions}
For the challenging case of implicit
questions, such as $q_3$ or $q_4$, the temporal value cannot be extracted from the question directly. 
To resolve this problem, 
we devise a novel mechanism,
the \textit{implicit question resolver},
based on recursively invoking the temporal QA system itself.
To this end, the implicit temporal constraint in the question is identified
and transformed into an \textit{intermediate question}.
For instance, the intermediate question
for $q_4$ would be
\phrase{when Freddie Mercury lead singer of Queen?}.
For $q_3$, 
the temporal value should be a time interval
(\struct{August\,1975\,-\,September\,1975}).
Thus, two intermediate questions are required:
(i) $q_{31}$: \phrase{When Queen recorded Bohemian Rhapsody start date?},
and (ii) $q_{32}$: \phrase{When Queen recorded Bohemian Rhapsody end date?}.
Although these formulations are ungrammatical, the QA system can process them properly, being robust to such inputs.

The intermediate questions
are fed into \faith
as a recursive call,
to obtain the explicit temporal value for filling
the {\tsf} of the original question.
The {\tsf} for $q_3$ thus becomes:
\begin{Snugshade}
$\langle$
    {question entity}: \phrase{Queen},\\
    \indent {question relation}: \phrase{recorded company},\\
    \indent expected answer type: \phrase{record company},\\
    \indent temporal signal: \struct{overlap},\\
    \indent temporal category: \struct{implicit},\\
    \indent temporal value: \struct{August\,1975\,-\,September\,1975}
$\rangle$
\end{Snugshade}
\noindent Note the similarity to the {\tsf}
of the explicit temporal question
$q_1$.

\myparagraph{Generating intermediate questions}
The intermediate questions are generated by a fine-tuned sequence-to-sequence (Seq2seq) model, specifically BART~\cite{lewis2020bart}.
A major obstacle, though, is that no prior dataset has suitable annotations, and collecting such data at scale is prohibitive.
Therefore, we generated training data using InstructGPT~\cite{ouyang2022training}, leveraging its
\textit{in-context learning}~\cite{brown2020language} capabilities.
We randomly select $8$ implicit questions from our train set and label them manually.
For each question, we give the intermediate question and the expected answer type as output.
The exact prompts used are shown in Table~\ref{tab:prompts} in the Appendix.

The expected answer type of an intermediate question can be \struct{date} or \struct{time interval}.
When the expected answer type is a time interval (e.g., for $q_3$),
two intermediate questions are created,
appending \phrase{start date} and \phrase{end date}
to the generated intermediate question,
respectively (see $q_{31}$ and $q_{32}$ as example).

We use this technique
to annotate all implicit questions in the train and dev sets, 
obtaining training data
for fine-tuning 
the BART model.
Note that 
GPT is used only for the generation of training data.
It is not used
at run-time to avoid its 
(computational, monetary, and environmental) costs and dependency on black-box models.

\myparagraph{Constructing the \tsf}
We also use a fine-tuned
Seq2seq model, again BART, for generating the values for the question entity, question relation, expected answer type, temporal signal, and temporal category slots of the {\tsf} representation.
The training data
this {\tsf} construction model is obtained via
(i) distant supervision (for question entity and question relation)~\cite{christmann2022conversational},
(ii) KB-type look-ups (for expected answer type),
and (iii) annotations in the benchmark (for temporal signal 
and temporal category).
Further detail in Sec.~\ref{sec:distant-supervision}.

The temporal values are obtained via the recursive
mechanism discussed above for implicit questions, and
via SUTime~\cite{chang2012sutime} and regular expression matching
for explicit questions. Phrases like \phrase{today} or \phrase{current} are considered as well and properly normalized.
We use the creation time of the question~\cite{campos2014survey}, as provided in the benchmarks, as reference time.

The TSF generated in this understanding stage is used for representing the
temporal information need in the subsequent retrieval and answering stages,
capturing its key temporal characteristics.

\vspace*{-0.1cm}
\subsection{Faithful Evidence Retrieval}
\label{subsec:fer}
In this stage, we first retrieve evidence from heterogeneous sources,
and then prune out information inconsistent with the temporal constraint
expressed by the temporal signal and value in the {\tsf}.

\myparagraph{Heterogeneous retrieval}
This step largely follows the general-purpose QA method of~\cite{christmann2022conversational},
and makes use of entity linking.
Entity mentions in the input are identified and linked via \clocq~\cite{christmann2022beyond}.
The input here is the concatenation of the \textit{question entity},  the \textit{question relation}, and the \textit{expected answer type} of the {\tsf}.
For the
resulting
linked
entities, we
retrieve the Wikipedia pages for extracting
text, tables, and infoboxes.
Further, KB-facts with the linked entities are obtained from Wikidata.

All retrieved pieces of evidence are \textit{verbalized}~\cite{oguz2021unikqa} into textual sentences, for uniform treatment.
The KB-facts are verbalized by concatenating their individual parts;
the text evidence is split into sentences;
table rows are transformed
by concatenating the individual $\langle$column headers, cell value$\rangle$ pairs;
infoboxes are handled
by linearizing all attribute-value pairs.

\myparagraph{Temporal pruning}
Explicit temporal expressions in the retrieved pieces of evidence are identified and normalized similarly as in the understanding stage.
Evidence that does not match the temporal criteria is pruned out. 
We address two kinds of 
situations:
\squishlist
\item[(i)] the question aims for a temporal value as answer and does not have any temporal constraints (e.g., \phrase{When \dots ?});
\item[(ii)] the question has a temporal constraint which needs to be matched by the evidence.
\squishend
In the first case, 
all evidence that does not contain any temporal values,
and is thus unable to provide the answer,
is dropped.
In the second case, we remove pieces of evidence that do not match the temporal constraint,
to ensure that answers are faithful to the temporal intent of the question.

The retrieval output
is a 
smaller set of evidence pieces, faithfully reflecting the temporal constraints of the question.
The final answer and its explanatory evidence are computed from this pool.

\subsection{Explainable Heterogeneous Answering}
\label{subsec:eha}
In the final stage, the answer is derived from this set of
evidence pieces that is already known to satisfy
the temporal conditions.

Since this part is not the main focus of this work,
we employ a state-of-the-art answering model for
general-purpose QA.
We use the answering stage of \explaignn~\cite{christmann2023explainable}
that is based on graph neural networks (GNNs),
and computes a subset of supporting evidence for the predicted answer.
Thus, we ensure that the answer can be traced back through
the entire system including the answering stage,
for end user explainability.
The input query to the GNNs
is the concatenation of the
question entity,
question relation,
and expected answer type.

%% file: sections/04-benchmark.tex
\begin{figure} [t]
     \includegraphics[width=\columnwidth]{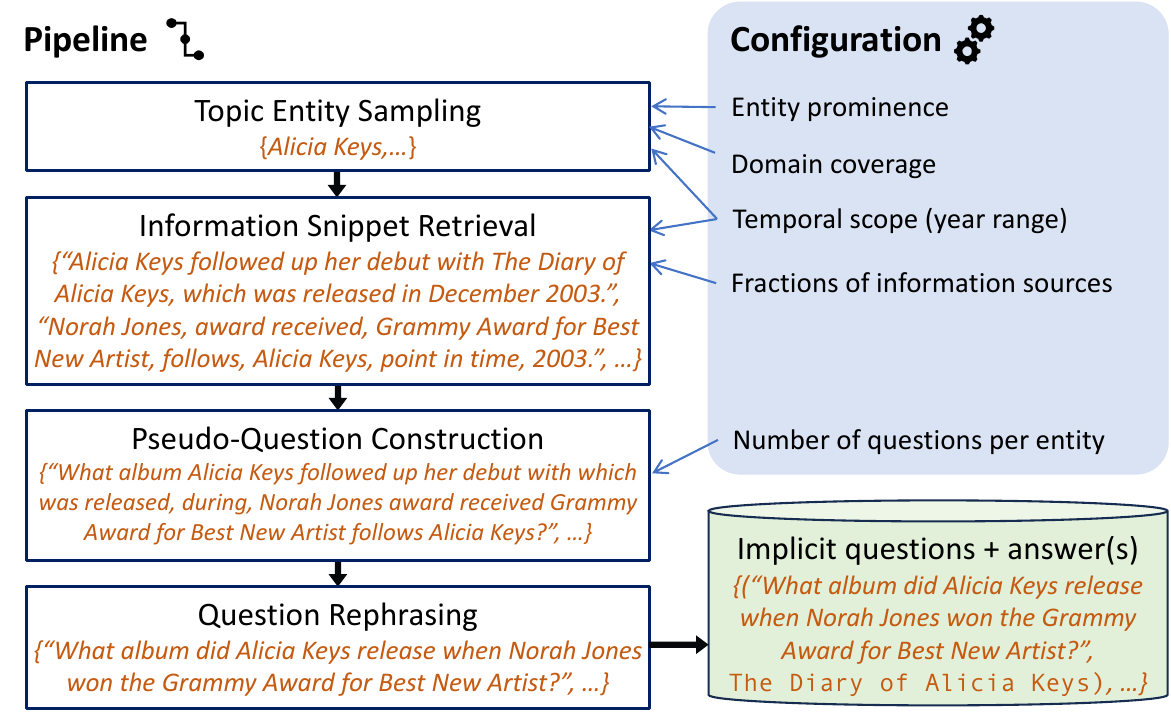}
     \vspace*{-0.5cm}
     \caption{Steps to create implicit questions with our
     proposed methodology, highlighting the key configurable parts.}
     \vspace*{-0.4cm}
     \label{fig:tiq-pipeline}
\end{figure}

\vspace*{-0.2cm}
\section{\benchmark benchmark}
\label{sec:benchmark}

Most existing benchmarks for temporal QA,
like
\textsc{TempQuestions}~\cite{jia2018tempquestions},
\timequestions~\cite{jia2021complex} or \textsc{TempQA-Wd}~\cite{neelam-etal-2022-sygma},
have only few implicit questions ({$209$, $1{,}476$, and $154$, respectively)}, 
falling short of evaluating one of the key challenges in temporal QA. \textsc{CronQuestions}~\cite{saxena2021question} \ZJadded{and \textsc{TEMPREASON}~\cite{tan2023towards} have} a larger fraction of implicit questions,
but these are based on a small set of hand-crafted rules.
Thus, the questions lack \textit{syntactic diversity}.
Further, 
questions in these benchmarks
are always answerable using a single information source
(either KB or text corpus).

Therefore, we 
construct a new benchmark with a primary focus on challenging and diverse implicit questions. 
The obvious idea of using crowdsourcing is expensive and error-prone.
Also, 
crowdworkers increasingly use LLMs
as a shortcut~\cite{veselovsky2023artificial}.
Thus, we pursue an automated process instead.
To ensure that questions are not specific to a single input source,
our process considers multiple sources:
Wikipedia text and infoboxes, and the Wikidata KB.

\vspace*{-0.2cm}
\subsection{Construction Methodology}

\myparagraph{Overview}
An implicit question
has two parts:
the \textit{main question} that specifies
the information need disregarding time
(e.g., \phrase{Queen's lead singer} for $q_4$),
and the \textit{implicit part} that provides the
temporal constraint (e.g., \phrase{after Freddie Mercury} for $q_4$).
The key idea is to build each of the two parts from independent pieces of evidence, denoted as \textit{information snippets}.
The two snippets can come from very different sources, but need to be thematically related.
%
This construction process operates as follows:
\squishlist
\item[(i)] sample a set of topic entities to start with;
\item[(ii)] retrieve temporal information snippets for each such topic entity from Wikipedia text, Wikipedia infoboxes, and Wikidata;
\item[(iii)] concatenate information snippets using a suitable temporal signal
and construct an interrogative sentence, a  \textit{pseudo-question};
\item[(iv)] rephrase the pseudo-question into a natural question using a generative model.
\squishend
An overview of this process is provided in Fig.~\ref{fig:tiq-pipeline},
including an example case of constructing an implicit question.
Naturally, implicit constraints are
global events (e.g., the COVID pandemic),
or major events for a specific entity
(e.g., a prestigious award).
%

\myparagraph{Sampling topic entities}
To obtain
significant events,
we start with Gregorian calendar year pages in Wikipedia
(e.g., {\small\url{https://en.wikipedia.org/wiki/2023}})
that list notable events.
From the pages for the years \struct{1801}\,-\,\struct{2025},
we collect information snippets about such significant events. 
The entities in these snippets constitute
the set of topic entities
(href anchors are used for entity linking~\cite{ferragina2010tagme}).

In our example in Fig.~\ref{fig:tiq-pipeline} this set includes
\struct{Alicia Keys}.



\myparagraph{Retrieving the grounding information snippets}
%
We collect snippets about notable events in these year pages,
and augment them with
salient information about the topic entity from
(i) the first five sentences ($\simeq$ first passage) of the entity's Wikipedia page, 
(ii)
the respective
Wikipedia infobox,
and (iii) the
Wikidata facts.

As candidates for the main question part, we consider all information snippets
that are retrieved for a topic entity from Wikipedia text, infoboxes and Wikidata, irrespective of their salience.
To avoid questions that are trivially answerable without considering the temporal condition, 
multiple candidate snippets are retrieved for the main question, with different temporal scopes (e.g., a band's singers from different epochs).
This is implemented by measuring semantic similarity among candidates using a SentenceTransformer\footnote{\url{https://huggingface.co/sentence-transformers/paraphrase-MiniLM-L6-v2}}~\cite{reimers-2019-sentence-bert}.


\myparagraph{Creating a pseudo-question}
Among the retrieved snippets for an entity,
we identify pairs of candidate snippets that can be connected by a temporal conjunction/preposition (\phrase{during}, \phrase{after} and \phrase{before}).
For such a pair, the temporal scopes have to be consistent with the temporal conjunction.
A valid pair
for the
conjunction \phrase{during} 
would be:
\phrase{Alicia Keys followed up her debut with The Diary of Alicia Keys, which was released in December 2003.} (main question part from Wikipedia text)
and
\phrase{Norah Jones, award received, Grammy Award for Best New Artist, follows, Alicia Keys, point in time, 2003.} (implicit part from KB).
A \textit{pseudo-question} is created by concatenating the main part with the
conjunction and the
implicit part.
The answer is an entity (not the topic entity) from the main part
(\struct{The Diary of Alicia Keys}).
The answer is substituted by the    
prefix \phrase{what} followed by the most frequent KB-type of the answer
(\struct{album} in this case).

The pseudo-question for the example is:
\phrase{What album Alicia Keys followed up her debut with which was released, during, Norah Jones award received Grammy Award for Best New Artist follows Alicia Keys?},
which is an ungrammatical and unnatural formulation.

\myparagraph{Rephrasing to a natural question}
Therefore, in the last step, we rephrase the pseudo-question
to a natural formulation.
We use InstructGPT~\cite{ouyang2022training}
with $8$ demonstration examples
(pseudo-questions and their natural re-phrasings),
to generate the final question\footnote{The prompt is given in Table~\ref{tab:prompt-tiq-rephrasing} in the Appendix.}.

The pseudo-question of the example is rephrased into the following implicit question:
\phrase{What album did Alicia Keys release when Norah Jones won the Grammy Award for Best New Artist?}


\vspace*{-0.3cm}
\subsection{Benchmark Characteristics}
\myparagraph{Topic entities}
For creating \benchmark
(\underline{T}emporal \underline{I}mplicit
\underline{Q}uestions) we started with the years \struct{1801}-\struct{2025} and obtained an initial set of {$229{,}318$} entities.
From this set, we uniformly sampled 
$10{,}000$
topic entities based on their frequency, to capture a similar amount of long-tail 
and more prominent entities (see Table~\ref{tab:statistics} for details). 
These fractions can be configured as required.
Since some entity types were over-represented in the calendar year pages
(e.g., politicians or countries), we also ensured that individual entity
types are not taking up more than $10\%$ of the topic entities.
In general, the topic entity set allows to control the domain coverage
within the generated implicit questions, by choosing
entities of the desired types.

\begin{figure} [t]
    \includegraphics[width=0.8\columnwidth]{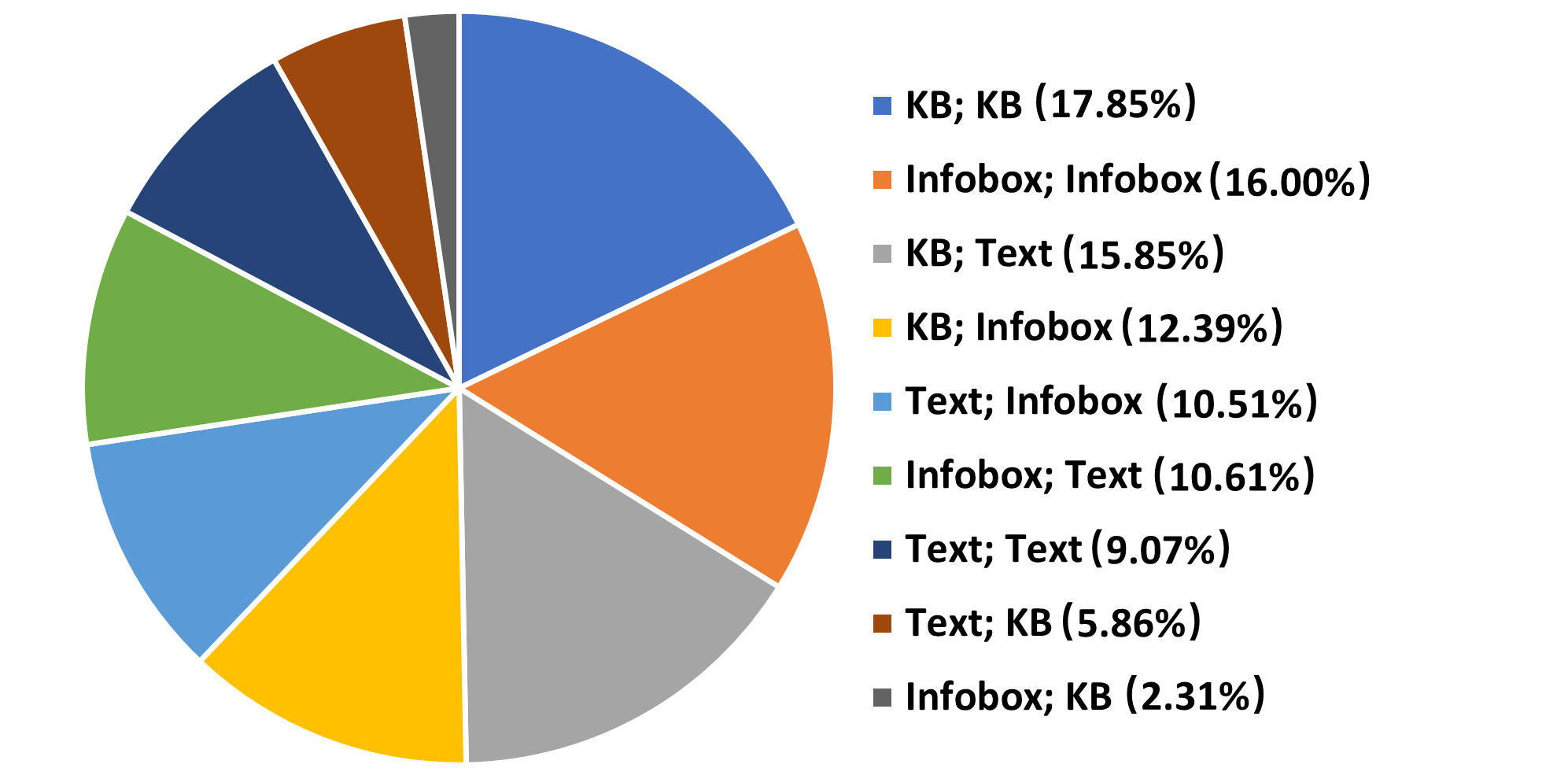}
     \vspace*{-0.3cm}
     \caption{Distribution of questions over input source combinations (source for main part ; source for implicit part).}
     \label{fig:benchmark-sources}
     \vspace*{-0.3cm}
\end{figure}

We did not
specifically
configure the proportions to which the individual
information sources are used within the questions,
since we observed a naturally diverse distribution.
Fig.~\ref{fig:benchmark-sources} 
shows the distribution among source combinations for initiating
the main and implicit part.
The questions are finally split into train ($6{,}000$), dev ($2{,}000$), and test sets ($2{,}000$).
Table~\ref{tab:statistics} shows the basic statistics,
and Table~\ref{tab:examples} shows representative questions of the \benchmark benchmark.

\myparagraph{Meta-data}
\benchmark provides implicit questions and gold answers, as strings as well as
canonicalized to Wikipedia and Wikidata.
The meta-data 
includes the information snippets grounding the question,
the sources these were obtained from,
the explicit temporal value expressed by the implicit constraint,
the topic entity, the question entities detected in the snippets,
and the temporal signal.

The \benchmark dataset is available at \textbf{\url{https://faith.mpi-inf.mpg.de}}.


\begin{table} [t] \small
	\centering
	\caption{Basic statistics for \benchmark.}
    \vspace*{-0.4cm}
	\resizebox*{\columnwidth}{!}{
        \begin{tabular}{p{4.1cm} p{4.6cm}}
            \toprule
            \textbf{Sources}    & Wikipedia text, infoboxes, and Wikidata   \\
            \textbf{Questions}  & 10,000 (train: $6{,}000$, dev: $2{,}000$, test: $2{,}000$)\\
            \midrule
            \textbf{Avg. question length}         & $17.96$ words   \\
            \textbf{Avg. no. of question entities}     & $2.45$\\
            \textbf{Unique topic entities covered}        & $10{,}000$ \\
            \textbf{Long-tail topic entities covered} & $2{,}542$ (with $<20$ KB-facts)\\
            \textbf{Prominent topic entities covered} & $2{,}613$ (with $>500$ KB-facts)\\ 
            \bottomrule
        \end{tabular}
    }	
	\label{tab:statistics}
    \vspace*{-0.3cm}
\end{table}

\begin{table*} [t] 
	\centering 
	\caption{Representative questions from the \benchmark benchmark.
    The sources below indicate the source that was used for populating the [main question part; implicit question part] of the implicit question.
 }
    \vspace{-0.4cm}
	\resizebox*{\textwidth}{!}{
	   \begin{tabular}{p{4cm} | p{4cm} | p{4.4cm} | p{4cm} | p{4cm}}
           \toprule
           1. \utterance{Who bought the Gainesville Sun after it was owned by Cowles Media Company?}
        &	2. 
        \utterance{During Colin Harvey's senior football career, which club was he a member of while he played for the England national football team?}
        &	3. \utterance{Which album released by Chris Brown topped the Billboard 200 when he was performing in Sydney?}	&		4. \utterance{What television series was Hulk Hogan starring in when he signed with World Championship Wrestling?}  &		5. \utterance{Who was Bristol Palin's partner before she participated in the fall season of Dancing with the Stars, and reached the finals, finishing in third place?} \\
        \struct{The New York Times Company}
        &	\struct{Everton F.C.}
        &	\struct{Fortune}		&	\struct{Thunder in Paradise} &	\struct{Levi Johnston} \\ \relax
			     [Text; KB] 	&	[Infobox; KB]	&		[Text; Infobox] 	&			 [Text; Text] &			 [Infobox; Text]	\\
           \midrule
         6. \utterance{During the onset of the COVID-19 pandemic, who was the New York City head of government?}	&
         7. \utterance{Who was the chief executive officer at Robert Bosch GmbH before revenue reached €78.74 billion?} &
         8. \utterance{After graduating from the Rostov-on-Don College of Economics and Finance, which political party did Gyula Horn join?} &	
         9. \utterance{Which national football team did Carlos Alberto Torres manage before joining Flamengo?} &
         10. \utterance{What university did Robert Lee Moore work for after Northwestern University?}\\
		  \struct{Bill de Blasio}	&	
          \struct{Volkmar Denner}
         &	\struct{Hungarian Working People's Party}
         &	\struct{Oman national football team} &	
         \struct{University of Pennsylvania}
         \\ \relax
			[KB; Text]	&		[KB; Infobox] 	&		[Infobox; Text] 	&		[Infobox; Infobox] &		[KB; KB]\\
	       \bottomrule
	   \end{tabular}
    }
	\label{tab:examples}
 \vspace*{-0.2cm}
\end{table*}

%% file: sections/06-results.tex
\begin{table} 
    \caption{
    Main results comparing the performance of \faith
    against baselines on the \textit{test} sets of \benchmark and \timequestions.} 
    \vspace*{-0.4cm}
    \newcolumntype{G}{>{\columncolor [gray] {0.90}}c}
    \newcolumntype{L}{>{}c}
      \resizebox*{\columnwidth}{!}{
    	\begin{tabular}{l G G G c c c}
    		\toprule
    		    \textbf{Benchmark} $\rightarrow$
                & \multicolumn{3}{G}{\textbf{\benchmark}} 
                & \multicolumn{3}{c}{\textbf{\timequestions}} \\

                \textbf{Method} $\downarrow$
                & \textbf{P@1} & \textbf{MRR} & \textbf{Hit@5} 
                & \textbf{P@1} & \textbf{MRR} & \textbf{Hit@5} \\
    	\midrule	    
            \textbf{\textsc{InstructGpt}~\cite{ouyang2022training}}
                &	$0.237$ &	n/a &	n/a
                &	$0.224$ &	n/a   &	n/a\\ 

                \textbf{\textsc{Gpt-4}~\cite{openai2023gpt4}}
                &	0.236 &	n/a &	n/a
                &	$0.306$  &	n/a   &	n/a\\

            \midrule
                \textbf{\uniqorn~\cite{pramanik2021uniqorn}}
                &	\ZJ{$0.236$} &	\ZJ{$0.255$} &	\ZJ{$0.277$}
                &	$0.331$  &	$0.409$   &	$0.538$ \\
    
                \textbf{\unikqa~\cite{oguz2021unikqa}}
                &	$0.425$ &	$0.480$ &	$0.540$
                &	$0.424$  &	$0.453$   &	$0.486$\\
        		\textbf{\explaignn~\cite{christmann2023explainable}}
                &	$0.446$ &	$0.584$ &	$0.765$
                &	$0.525$  &	$0.587$   &	$0.673$\\

            \midrule
                \textbf{\tempoqr~\cite{mavromatis2022tempoqr}}
                &	$0.011$  &	$0.018$   &	$0.022$
                &	$0.438$  &	$0.465$   &	$0.488$\\

                 \textbf{\cronkgqa~\cite{saxena2021question}}
                &	$0.006$  &	$0.011$   &	$0.014$
                &	$0.395$  &	$0.423$   &	$0.450$\\

                \textbf{\exaqt~\cite{jia2021complex}}
                &	$0.232$  &	$0.378$   &	$0.587$
                &	$0.565$  &	$0.599$   &	$0.664$\\

            \midrule
                \textbf{\temporalqa (Proposed)}
                &	\ZJ{$0.491$}  &	\ZJ{$0.603$}   &	\ZJ{$0.752$}
                &	\ZJ{$0.535$}  &	\ZJ{$0.582$}   &	\ZJ{$0.635$}\\
               
                \textbf{Un-\temporalqa}
               &\ZJ{${0.459}$}  &\ZJ{${0.604}$}& \ZJ{${0.799}$}
                &\ZJ{${0.571}$}  &	\ZJ{${0.640}$}   &	\ZJ{${0.724}$}\\
                
    		\bottomrule
    	\end{tabular} 
    }
    \label{tab:main-res}
    \vspace*{-0.2cm}
\end{table}

\section{Experiments}
\label{sec:exps}

\subsection{Experimental Setup}
\label{sec:setup}

\myparagraph{Benchmarks}
We conduct experiments on our new \benchmark benchmark
and \timequestions~\cite{jia2021complex}, which has been actively used in recent work on temporal QA.
For ordinal questions
(e.g., \phrase{what was the first album by Queen?}) in \timequestions, we apply the same method as outlined in
Sec.~\ref{sec:method}, without applying any temporal filtering. 




\myparagraph{Metrics}
We use the standard QA metrics
precision at 1 ({P@1}),
mean reciprocal rank ({MRR}),
and hit at 5 ({Hit@5})~\cite{saharoy2022question}.

\myparagraph{Baselines}
We compare \faith with a suite of baselines,
covering a diverse
range of competitors:
\squishlist
    \item \textbf{Generative LLMs}.
        We compare with \textbf{\textsc{InstructGpt}}~\cite{ouyang2022training} (``text-davinci-003'')
        and \textbf{\textsc{Gpt-4}}~\cite{openai2023gpt4} (``gpt-4'') using the OpenAI API\footnote{\url{https://platform.openai.com}}.
        We tried different prompts, and found the following to perform best:
        \phrase{Please answer the following question by providing the crisp answer entity, date, year, or number.}.
        For computing P@1,
        we check whether the generated answer string matches
        with the label or any alias of the gold answer. If this is the case,
        P@1 is $1$, else $0$. Other (ranking) metrics are not applicable for LLMs.

    \item \textbf{Heterogeneous QA methods}.
        Further, we compare against a range of recent general-purpose methods
        for heterogeneous QA:
        \textbf{\textsc{Uniqorn}}~\cite{pramanik2021uniqorn},
        \textbf{\textsc{UniK-Qa}}~\cite{oguz2021unikqa},
        and the vanilla \textbf{\textsc{Explaignn}}~\cite{christmann2023explainable}.

    \item \textbf{Temporal QA methods}.
        We also compare with state-of-the-art methods for temporal QA:
        \textbf{\textsc{TempoQR}} (TempoQR-Hard)~\cite{mavromatis2022tempoqr},
        \textbf{\textsc{CronKGQA}}~\cite{saxena2021question},
        and \textbf{\textsc{Exaqt}}~\cite{jia2021complex}.
  \squishend      

\noindent Finally, we show results for a variant of our approach, which does \textit{not prune out} evidence temporally-inconsistent with the temporal constraint, i.e. drops the temporal pruning component.
We term this variant \textbf{Un-\faith}.


\vspace*{0.1cm}
\myparagraph{Configuration}
Wikidata~\cite{vrandevcic2014wikidata} is used as the KB for \faith and all baselines.
We use Wikipedia text, tables and infoboxes as additional information sources
for methods operating over heterogeneous sources.
The BART models are
initialized
via Hugging Face\footnote{\url{https://huggingface.co}}. 
We use AdamW as optimizer with a learning rate of $5\times$$10^{-5}$, batch size of $10$, weight decay of $0.01$, $5$ epochs, and $500$ warm-up steps.
\explaignn is run using the public code\footnote{\url{https://github.com/PhilippChr/EXPLAIGNN}}, retaining the original settings and parameters for optimization.

For \faith, we choose the candidate at rank $1$ as the answer for intermediate questions in the implicit question resolver.
In case too many evidences are obtained as input to the answering stage,
we consider 
the top-$100$ evidences as computed by a BERT-based re-ranker~\cite{nogueira2019passage}.
Further detail is given in the Appendix~\ref{sec:scoring-model}. We follow an epoch-wise evaluation strategy for each module and baseline,
and take the version with the best performance on the respective dev set.
All training processes and experiments are run on a single GPU (NVIDIA Quadro RTX 8000, 48 GB GDDR6).

\vspace*{-0.1cm}
\subsection{Main Results}
\label{sec:main-res}
Answering performance of \faith and baselines on \timequestions and on \benchmark are in Table~\ref{tab:main-res}.

\myparagraph{\faith outperforms baselines on \tiq} 
The main insight from Table~\ref{tab:main-res} is that \faith
surpasses all baselines on the \tiq dataset for P@1,
which is the most relevant metric,
demonstrating the benefits of our proposed method for answering implicit temporal questions.
Temporal QA methods operating over KBs lack the required coverage on the \tiq dataset,
and perform worse than general-purpose QA methods operating over heterogeneous sources.
\explaignn comes close to the performance of \faith,
and even slightly improves on the 
Hit@5 metrics.
Note, however, that \explaignn and all other baselines
do not verify that temporal constraints are met during answering.
Thus, the most prominent among answer candidates may simply be picked up,
even if no temporal information is provided or matching.
Such possibly ``accidental'' and \textit{unfaithful}
answers are, by design, not considered by \faith.

\myparagraph{Trade-off between faithfulness and answering performance}
Results for Un-\faith illustrate the effect of this phenomenon on our approach:
especially the MRR and Hit@5 results are substantially improved.
Consequently, Un-\faith outperforms all competitors on \timequestions.
However, its answers are not always faithfully grounded in evidence sources.
These results emphasize the trade-off between faithfulness and answering performance.


\myparagraph{\faith shows robust performance on \timequestions} 
\faith also shows strong performance on the \timequestions benchmark, on which it outperforms
all baselines on P@1, except for \exaqt. This indicates the robustness of \faith across different datasets.
Existing methods for temporal QA show major performance gaps between the two benchmarks:
the P@1 of the strongest method on \timequestions, \exaqt, substantially drops
from $0.565$ at P@1 to $0.232$ on the \benchmark benchmark.
Note that all methods are trained on the specific benchmark, if applicable.


\myparagraph{LLMs fall short on temporal questions} 
Another key insight from Table~\ref{tab:main-res} is that current LLMs are clearly not capable
of answering temporal questions.
\textsc{InstructGpt} and \textsc{Gpt-4} can merely answer $\simeq23$-$30$\% of the questions correctly,
and are constantly underperforming \faith and baselines operating over heterogeneous sources.
One explanation is that reasoning with continuous variables, such as time, is a well-known weakness of LLMs~\cite{dhingra2022time}.



\begin{table} \small
    \caption{
    Comparing the faithfulness of \faith and Un-\faith for correct answers, and how often temporal constraints are violated or ignored.
    } 
    \vspace*{-0.4cm}
    \newcolumntype{G}{>{\columncolor [gray] {0.90}}c}
    \newcolumntype{L}{>{}c}
      \resizebox*{\columnwidth}{!}{
    	\begin{tabular}{l G G c c}
    		\toprule
    		    \textbf{Benchmark} $\rightarrow$
               & \multicolumn{2}{G}{\textbf{\benchmark}} &
               \multicolumn{2}{c} {\textbf{\timequestions}} \\
               
               \midrule

                 & & \textbf{Temporally}
                 & & \textbf{Temporally}\\

                \textbf{Method} $\downarrow$
                 & \textbf{Faithful} & \textbf{Unfaithful}
                 & \textbf{Faithful} & \textbf{Unfaithful} \\
    		    
            \midrule
        		\textbf{\faith}
                &	$0.95$  &  $0.00$
                &	$0.94$  & $0.01$\\

                \textbf{Un-\faith}
                 &	$0.90$ & $0.08$
                  &	$0.87$ & $0.13$\\
    		\bottomrule
    	\end{tabular} 
    }
    \label{tab:faithfulness-evaluation}
    \vspace{-0.2cm}
\end{table}

\subsection{Faithfulness Evaluation}
\label{sec:faith-unfaith-analysis}

Our main results in Table~\ref{tab:main-res}
indicate that ignoring the temporal condition of the question
can yield improvements on automatic metrics
(compare performance of \faith vs. Un-\faith on \timequestions).
However, we observe that this can lead to critical failure cases of QA systems and
sometimes boils down to lucky guesses of the answer based on priors (e.g., prominence of an answer candidate).

\myparagraph{\faith refrains to answer in absence of consistent evidence}
If there is no temporal information associated with the evidence of candidate answers, or the temporal information does not satisfy the temporal constraint, \faith will refuse answering the question.
For example, for the question \phrase{Who did Lady Jane Grey marry on the 25th of May 1533?}, there is no answer satisfying the temporal constraint because \utterance{Lady Jane Grey} did not marry anyone \utterance{on the 25th of May 1533}, since she was only born
four years
later in 1937.
However, all of the baselines provide an answer to the question, without indicating that the temporal constraint is violated.

Since
questions without a temporally-consistent answer are not available at large scale,
we randomly sample $500$ explicit questions from \timequestions, and replace the temporal
value with a random date (e.g., \phrase{12 October 6267}).
None of the resulting questions has a temporally-consistent answer.
As expected, the competitors still provide answers\footnote{Except for the LLMs for which we are not able to investigate the behavior at scale, since they would often generate longer texts.}.
In contrast, \faith successfully refrained from answering for $467$ of the $500$ questions ($93.4\%$).
Upon investigating the failure cases, we noticed that the date recognition identifies
four-digit numbers as years matching with the constraint (e.g., in the infobox entry \phrase{Veysonnaz, SFOS number, 6267}).

\myparagraph{Fallback to Un-\faith}
Completely refraining from answering
could also be sub-optimal:
the user might have made a typo
(e.g., \phrase{May 1533} instead of \phrase{May 1553}).
We investigated to fall back to Un-\faith in such scenarios, which could be indicated to end users with an appropriate warning.
Performance on both datasets was slightly improved: the P@1 metric increased from $0.491$ to $0.492$ on \benchmark and from $0.535$ to $0.539$ on \timequestions.
We further investigated to fall back to Un-\faith in case \faith answered incorrectly.
The P@1 metric was improved substantially on both datasets:
from 
{$0.491$ to $0.622$}
on \benchmark and from {$0.535$ to $0.653$}
on \timequestions.

{\myparagraph{Manual analysis}}
Finally, we investigated the faithfulness of \textit{correct} answers provided by \faith and Un-\faith, to understand
how often the question is answered correctly even though the evidence is not faithful to the question.
To analyze this qualitatively, we randomly selected $100$ questions (from each benchmark) for which both \faith and Un-\faith answered correctly,
and then manually verified the faithfulness,
based on the definition in Sec.~\ref{sec:concepts}.
Results are in Table~\ref{tab:faithfulness-evaluation}.
\faith provides faithful answers and evidence in $95\%$/$94\%$ of the time.
By design, answers are faithful to the temporal constraints in the question {(except for one question which specifies two different temporal constraints)}.
In comparison, Un-\faith violates or ignores the temporal condition in $8\%$/$13\%$ of the cases.

For example, to answer the question \phrase{What movies starring Taylor Lautner in 2011?} (answer: \struct{Abduction}),
the evidence for \faith is \phrase{Taylor Lautner, Year is 2011, Title is Abduction, Role is Nathan Harper} (from table),
while the evidence for Un-\faith is \phrase{Abduction, cast member, Taylor Lautner}  (from KB).
Even though both pieces of evidence mention the correct answer \struct{Abduction}, Un-\faith fails to satisfy the temporal constraint (\phrase{in 2011}) with its evidence.

\begin{table} 
    \caption{Ablation study using different source combinations as input for \temporalqa on \textit{dev} sets. Note that \temporalqa is trained using \textit{all sources}
    as input for all cases.} 
    \vspace*{-0.4cm}
    \newcolumntype{G}{>{\columncolor [gray] {0.90}}c}
    \newcolumntype{L}{>{}c}
      \resizebox*{\columnwidth}{!}{
    	\begin{tabular}{l G G G c c c}
    		\toprule
    		    \textbf{Benchmark} $\rightarrow$
                & \multicolumn{3}{G}{\textbf{\benchmark}} 
                & \multicolumn{3}{c}{\textbf{\timequestions}} \\

                \textbf{Method} $\downarrow$
                & \textbf{P@1} & \textbf{MRR} & \textbf{Hit@5} 
                & \textbf{P@1} & \textbf{MRR} & \textbf{Hit@5} \\
    		    
            \midrule
        		\textbf{KB}
                &	$0.293$  &	$0.368$   &	$0.468$
                &	{$0.425$}  &	{$0.464$}   &	{$0.513$}\\

                \textbf{Text}
                 &	$0.194$  &	$0.262$   &	$0.351$
                 &	$0.224$  &	$0.269$   &	$0.320$
                \\

                \textbf{Infoboxes}
                &	$0.169$  &	$0.223$   &	$0.296$ 
                
                &	$0.093$  &	$0.117$   &	$0.149$ 
                 \\

                \textbf{Tables}
               &	$0.032$  &	$0.057$   &	$0.083$  
                &	$0.078$  &	$0.094$   &	$0.114$
                \\
            
            \midrule
                \textbf{KB+Text}
                &	$0.429$  &	$0.527$   &	$0.649$
                &	$0.520$  &	$0.567$   &	$0.626$
                 \\    

                \textbf{KB+Tables}
                 &	 $0.299$  &	$0.379$   &	$0.480$
                 &	$0.435$ &	$0.479$   &	$0.536$
                \\     

                \textbf{KB+Infoboxes}
                &	 $0.384$  &	$0.488$   &	$0.634$ 
                &	$0.443$  &	$0.487$   &	$0.543$
                \\     

                \textbf{Text+Tables}
                &	 $0.196$  &	$0.267$   &	$0.362$ 
                &	$0.252$  &	$0.298$   &	$0.350$
                \\     

                \textbf{Text+Infoboxes}
                &	$0.283$  &	$0.372$   &	$0.490$ 
                &$0.251$  &	$0.299$   &	$0.355$
                
                \\     

                \textbf{Tables+Infoboxes}
               &	$0.179$  &	$0.244$   &	$0.331$
               &	$0.143$  &	$0.174$   &	$0.208$
                 \\   

            \midrule
                \textbf{All sources}
                &	$0.497$ &	$0.610$  &	$0.756$
               &	$0.538$  &	$0.583$   &	$0.639$
                 \\    
                
    		\bottomrule
    	\end{tabular} 
    }
    \label{tab:sources}
    \vspace{-0.2cm}
\end{table}

\begin{table} \small
    \caption{Ablation studies of \temporalqa on dev sets.} 
    \vspace*{-0.4cm}
    \newcolumntype{G}{>{\columncolor [gray] {0.90}}c}
    \newcolumntype{L}{>{}c}
    	\begin{tabular}{l G c}
    		\toprule
    		    \textbf{Benchmark} $\rightarrow$
                & \multicolumn{1}{G}{\textbf{\benchmark}} 
                & \multicolumn{1}{c}{\textbf{\timequestions}} \\

                \textbf{Method} $\downarrow$
                & \textbf{P@1} & \textbf{P@1}\\ 
            \midrule
        		\textbf{\temporalqa}
               &\ZJ{$0.497$}	
               &	\ZJ{$0.538$} \\ 
                \midrule

                \textbf{w/o temporal pruning}
                & \ZJ{$0.443$}	
                &	\ZJ{$0.573$} \\ 



                \textbf{w/o implicit question resolver}
               &	\ZJ{$0.467$} &	\ZJ{$0.559$}\\


               
                
                
                \textbf{w/o GNN-based answering} 
                 &	\ZJ{$0.316$} &	\ZJ{$0.399$}\\ 
    		\bottomrule
    	\end{tabular} 
    
    \label{tab:ablations}
    \vspace*{-0.3cm}
\end{table}

\begin{table*} [t] \footnotesize
\caption{Anecdotal examples that \faith answered correctly in \benchmark and \timequestions. \textbf{Evidence} shows the supporting information snippets along with their source provided in brackets. The part mentioning the predicted answer is in \textbf{bold}, and the detected temporal values are underlined. For the first example from the \benchmark benchmark, we show the answering process of the intermediate question, which can be used by end users to verify the entire answer derivation of the system.}
\centering
    \vspace*{-0.3cm}
    \resizebox*{\textwidth}{!}{
    	\begin{tabular}{l | p{14cm}}
    	\toprule
            \rowcolor{lightgray} \textbf{Benchmark} &\textbf{\benchmark} 	\\  
        \midrule
    
            \textbf{Question}	    & \textit{After managing FC Nantes, which football club did Antoine Raab take on next?}              \\
            \textbf{Answer}	    & \textbf{Stade Lavallois}              \\
            \textbf{TSF}	 & \cellcolor[rgb]{0.9255,0.9255,0.9255}$\langle$ {question entity}: \phrase{Antoine Raab, FC Nantes football}, question relation: \phrase{After managing which club did take on next}, expected answer type: \phrase{association football club}, temp. signal: \texttt{after}, temp. category: \texttt{implicit}, temp. value: \texttt{[1946, 1949]} $\rangle$ \\
            \textbf{Evidence}	     &\textit{Antoine Raab, Managerial career, \underline{1949–1950}, \textbf{Stade Lavallois}}. \textcolor{blue}{\textsc{(from Infobox)}} \\
            
        \midrule
        
            \textbf{Intermediate questions}	& {(i)} \textit{When Antoine Raab managed FC Nantes start date?}  \\
            &{(ii)} \textit{When Antoine Raab managed FC Nantes end date?}\\
             \textbf{Answers (to int. questions)}	& (i) \textbf{1946}, (ii) \textbf{1949} \\
          \textbf{TSFs (for int. questions)} &  \cellcolor[rgb]{0.9255,0.9255,0.9255}(i) $\langle$ {question entity}: \phrase{FC Nantes, start, Antoine Raab}, question relation: \phrase{When managed date}, expected answer type: \phrase{year}, temp. signal: \texttt{\_}; temp. category: \texttt{non-implicit}; temp. value: \texttt{\_} $\rangle$ \\
          & \cellcolor[rgb]{0.9255,0.9255,0.9255}(ii) $\langle$ {question entity}: \phrase{FC Nantes, end, Antoine Raab}, question relation: \phrase{When managed date}, expected answer type: \phrase{year}, temp. signal: \texttt{\_}; temp. category: \texttt{non-implicit}; temp. value: \texttt{\_} $\rangle$   \\
          \textbf{Evidence (for int. questions)}	     &(i, ii) \textit{Antoine Raab, Managerial career, \textbf{1946}–\textbf{1949}, FC Nantes}. \textcolor{blue}{\textsc{(from Infobox)}} \\
          &(ii) \textit{Antoine Raab, After the liberation of Nantes in 1944 Raab joined FC Nantes and played for the club until \textbf{1949}.} \textcolor{blue}{\textsc{(from Text)}} \\

        \midrule
     
            \rowcolor{lightgray} \textbf{Benchmark} &\textbf{\timequestions} 	\\
        \midrule
          \textbf{Question}  & \textit{What award did Thomas Keneally receive in the year 1982?} \\
          \textbf{Answer}  & \textbf{Booker Prize}   \\
    
           \textbf{TSF} & \cellcolor[rgb]{0.9255,0.9255,0.9255}{$\langle$ {question entity}: \phrase{Thomas Keneally}, question relation: \phrase{What award did receive in the year 1982}, expected answer type: \phrase{science award}, temp. signal: \texttt{overlap}, temp. category: \texttt{non-implicit}, temp. value: \texttt{1982} $\rangle$}\\
      
        \textbf{Evidence} 
            &\textit{Man \textbf{Booker Prize}, winner, Thomas Keneally, point in time, \underline{1982}, for work, Schindler's Ark}. \textcolor{blue}{\textsc{(from KB)}} \\

            &\textit{Thomas Keneally, Awards is \textbf{Booker Prize}, is Schindler's Ark, winner \underline{1982}}. \textcolor{blue}{\textsc{(from table)}} \\
            &\textit{Thomas Keneally, He is best known for his non-fiction novel Schindler's Ark, the story of Oskar Schindler's rescue of Jews during the Holocaust, which won the \textbf{Booker Prize} in \underline{1982}.} \textcolor{blue}{\textsc{(from Text)}} \\
            \bottomrule
	   \end{tabular}
    }
\label{tab:anecdotes}
\end{table*}

\subsection{In-depth Analysis}
\label{sec:analysis}

\myparagraph{Integrating heterogeneous sources is decisive}
We further
investigated the effect of integrating heterogeneous sources into \faith,
and tested giving each individual
source independently, and their pairwise combinations as input,
in comparison to the default setting with "All sources".
Results are in Table~\ref{tab:sources}.
Each information source contributes to the performance of \faith,
and integrating more information sources consistently enhances all metrics.



\myparagraph{Ablation studies}
We tested
variations of our pipeline on the dev sets.
Table~\ref{tab:ablations} shows
results for Un-\faith (w/o temporal pruning), results without the implicit time resolver,
and results with a Seq2seq model for answering (we used BART)
instead of the GNN-based approach.
Using a GNN-based answering approach plays a crucial role, and enhances not only
answering performance, but also explainability.
The implicit question resolver is decisive on \benchmark, but slightly decreases performance on \timequestions.
Un-\faith also shows strong performance on the dev sets.
However, all modules contribute to the explainability
and faithfulness of our approach.

\myparagraph{Anecdotal examples}
Table~\ref{tab:anecdotes} shows sample cases for which \faith provided the correct answer, and illustrates the answer derivation process providing traceable evidence for end users.

\myparagraph{Error analysis}
To better understand failure cases,
we conducted an error analysis measuring
the \textit{answer presence}
(i.e. whether the gold answer is among answer candidates)
throughout the pipeline.
We identified the following error cases and list their percentage among all failure cases
for \benchmark and \timequestions, respectively:
{(i) the answer was not found in the initial retrieval stage
($3.14$/$29.89$),
(ii) the answer is lost during temporal pruning
($22.00$/$25.81$),
(iii) the answer is lost during scoring/graph shrinking
($8.45$/$10.33$),
(iv) the answer is not considered among top-$5$ answers
($15.13$/$12.47$),
(v) the answer is among top candidates but not at rank $1$
($51.28$/$21.51$).}



%% file: sections/07-related.tex
\vspace*{-0.2cm}
\section{Related Work}
\label{sec:relatedwork}

\myparagraph{General-purpose QA}
Question answering has extensive work using
single sources
like KBs (e.g.,~\cite{yahya2012natural, yih2015semantic, berant2013semantic})
or text (e.g.,~\cite{chen2017reading, rajpurkar2016squad, izacard2021leveraging}).
%
Some works have shown
that integrating different 
sources
can
substantially improve
performance 
~\cite{ferrucci2012introduction,xu2016hybrid,xu2016question,savenkov2016knowledge,sun2018open,sun2019pullnet,chen2021open}.
\unikqa~\cite{oguz2021unikqa} 
verbalizes 
snippets from a KB, text, tables and infoboxes,
as input to a 
Fusion-in-decoder (FiD) model~\cite{izacard2021leveraging}
for answer generation.
\textsc{Udt-QA}~\cite{ma2021open} 
improved the verbalization technique.
\explaignn~\cite{christmann2023explainable}
constructs graphs among such verbalized snippets,
and applies graph neural networks
for computing answers and explanatory evidence.
None of these methods is geared for
temporal questions.

Another direction is to directly apply 
large language models (LLMs)
for QA
~\cite{brown2020language, raffel2020exploring, petroni2019language, devlin2019bert}.
However, LLMs cannot present traceable provenance for the generated outputs,
falling short on
faithfulness and explainability 
~\cite{adlakha2023evaluating, min2023factscore, mallen2023not}.
Also, LLMs struggle with reasoning on temporal conditions~\cite{dhingra2022time}.

\myparagraph{Temporal QA}
Prior work
that specifically targets temporal QA
~\cite{wu2020introducing, jia18tequila,jia2021complex,jiao2022improving,saxena2021question,xiao2022modeling,chen2022temporal,ding2022semantic,neelam-etal-2022-sygma,sharma2022twirgcn,mavromatis2022tempoqr,shang2022improving,yao2022terqa,chen2021dataset,long2022complex},
can
largely be divided into work using a KB
(e.g.,~\cite{jia2021complex, neelam-etal-2022-sygma, mavromatis2022tempoqr}),
and work using text 
(e.g.,~\cite{chen2021dataset, ning2020torque}).
\textit{Methods operating over KBs}, 
include template-based
~\cite{jia18tequila,ding2022semantic,neelam-etal-2022-sygma},
KB-embedding-based
~\cite{saxena2021question,xiao2022modeling,chen2022temporal,mavromatis2022tempoqr},
and graph-based methods~\cite{jia2021complex,sharma2022twirgcn,yao2022terqa}.
\textit{Methods using textual inputs} typically
involve an extractive or generative reader~\cite{chen2021dataset, ning2020torque}.


The three methods~\cite{jia2021complex,saxena2021question,mavromatis2022tempoqr} 
represent the state-of-the-art
on temporal QA. However, temporal constraints are handled solely
in the latent space, without explicitly (or \textit{faithfully})
pruning out temporally inconsistent answer candidates.
Other approaches are based on handcrafted rules, and hence 
bound to fail for unseen question patterns (e.g.,~\cite{jia18tequila}).
None of the existing work on temporal QA has considered incorporating heterogeneous sources.


\myparagraph{Temporal KBs}
There is
substantial work on temporal KBs 
~\cite{leblay2018deriving, garcia2018learning, cai2022temporal,wu2020temp, xu2021temporal, messner2022temporal, park2022evokg},
to
assign temporal scopes to KB facts.
Advances 
on the KB itself benefits QA, but is an orthogonal direction.

%% file: sections/08-conclusion.tex
\section{Conclusion}
\label{sec:conclusion}
This work 
targets complex temporal QA, and 
proposes a new approach for \textit{faithfully} answering
temporal questions, with focus on the 
challenging case of implicit temporal constraints.
Experiments show that our method \faith outperforms
the best unfaithful competitor on such implicit questions.
On other temporal questions, our method performs almost on par,
but adds the benefit of reliably matching the temporal conditions.
Faithfulness is an important element in enhancing the trustworthiness of QA systems.

\vspace*{0.2cm}
\myparagraph{Acknowledgements}
\ZJadded{We thank Rishiraj Saha Roy and Magdalena Kaiser from the Max Planck Institute for Informatics for useful inputs at various stages of this work. Zhen Jia was supported by NSFC (Grant No.62276215 and No.62272398).}

%% file: sections/10-appendix.tex
\begin{table} [t] \footnotesize
	\centering
    \caption{Prompt including demonstrations for rephrasing the pseudo-questions into natural questions.}
    \vspace*{-0.3cm}
	\resizebox*{\columnwidth}{!}{
        \begin{tabular}{p{11cm}}
            \toprule
            Please rephrase the following input question into a more natural question.\\\\
        
            \textit{Input}: What album Sting ( musician ) was released, during, Sting award received German Radio Award?\\
            \textit{Question}: which album was released by Sting when he won the German Radio Award?\\\\
            
            \textit{Input}: What human President of Bolivia was the second and most recent female president, after, president of Bolivia officeholder Evo Morales?\\
            \textit{Question}: Which female president succeeded Evo Morales in Bolivia?\\\\
            
            \textit{Input}: What lake David Bowie He moved to Switzerland purchasing a chalet in the hills to the north of , during, David Bowie spouse Angela Bowie?\\
            \textit{Question}: Close to which lake did David Bowie buy a chalet while he was married to Angela Bowie?\\\\
            
            \textit{Input}: What human Robert Motherwell spouse, during, Robert Motherwell He also edited Paalen 's collected essays Form and Sense as the first issue of\\Problems of Contemporary Art?\\
            \textit{Question}: Who was Robert Motherwell's wife when he edited Paalen's collected essays Form and Sense?\\\\
            
            \textit{Input}: What historical country Independent State of Croatia the NDH government signed an agreement with which demarcated their borders, during,\\Independent State of Croatia?\\
            \textit{Question}: At the time of the Independent State of Croatia, which country signed an agreement with the NDH government to demarcate their borders?\\\\
            
            \textit{Input}: What U-boat flotilla German submarine U-559 part of, before, German submarine U-559 She moved to the 29th U-boat Flotilla?\\
            \textit{Question}: Which U-boat flotilla did the German submarine U-559 belong to before being transferred to the 29th U-boat Flotilla?\\\\
            
            \textit{Input}: What human UEFA chairperson, during, UEFA chairperson Sandor Barcs?\\
            \textit{Question}: Who was the UEFA chairperson after Sandor Barcs?\\\\
            
            \textit{Input}: What human Netherlands head of government, during, Netherlands head of state Juliana of the Netherlands?\\
            \textit{Question}: During Juliana of the Netherlands' time as queen, who was the prime minister in the Netherlands?\\
            \bottomrule
        \end{tabular}
    }
	\label{tab:prompt-tiq-rephrasing}
\end{table}

\begin{table*} [t] \footnotesize
	\centering
	\caption{Prompts used to obtain the training data for generating intermediate questions, leveraging in-context learning.}
    \vspace*{-0.3cm}
	\resizebox*{\textwidth}{!}{
        \begin{tabular}{p{10cm} | p{13cm}}
            \toprule
             \textbf{\timequestions} & \textbf{\benchmark} \\
            \midrule
            Generate an explicit question and answer type for the implicit part of the temporal input question.
            & Generate an explicit question and answer type for the implicit part of the temporal input question.\\\\
    
            \textit{Input}: what position did djuanda kartawidjaja take after he was replaced by sukarano 
                \newline \textit{Output}: when djuanda kartawidjaja replaced by sukarano||date
            & \textit{Input}: Who was the second director of the Isabella Stewart Gardner Museum when it was built
                \newline \textit{Output}: When Isabella Stewart Gardner Museum was built||time interval\\\\
    
            \textit{Input}: american naval leader during the world war 2
                \newline \textit{Output}: when world war 2||time interval
            & \textit{Input}: When Wendy Doniger was president of the Association for Asian Studies, what publishing house was she based in New York
                \newline \textit{Output}: When Wendy Doniger was president of the Association for Asian Studies||time interval\\\\
    
            \textit{Input}: who became president after harding died
                \newline \textit{Output}: when harding died||date
            & {\textit{Input}: What administrative entity was Ezhou in before Huangzhou District became part of it
                \newline \textit{Output}: When Huangzhou District became part of Ezhou||date}\\\\
    
            \textit{Input}: who did luis suarez play for before liverpool
                \newline \textit{Output}: when luis suarez play for liverpool||time interval 
            & \textit{Input}: After Bud Yorkin became the producer of NBC's The Tony Martin Show, who was his spouse?
                \newline \textit{Output}: When Bud Yorkin became the producer of NBC's The Tony Martin Show||date\\\\
    
            \textit{Input}: which countries were located within the soviet union prior to its dissolution
                \newline \textit{Output}: when soviet union dissolution||date 
            & \textit{Input}: What book did Ira Levin write that was adapted into a film during the same time he wrote the play Deathtrap
                \newline \textit{Output}: When Ira Levin wrote the play Deathtrap||date\\\\
    
            \textit{Input}: who started the presidency earliest and served as president during wwii in the US 
                \newline \textit{Output}: when wwii||time interval
            & \textit{Input}: What basketball team was Nathaniel Clifton playing for when his career history with the Rens began
                \newline \textit{Output}: When Nathaniel Clifton's career history with the Rens began||time interval\\\\
    
            \textit{Input}: who replaced aldo moro as the minister of foreign affairs
                \newline \textit{Output}: when aldo moro replaced as minister of foreign affairs||date
            & \textit{Input}: What team did Stevica Ristić play for before joining Shonan Bellmare?
                \newline \textit{Output}: When Stevica Ristić joining Shonan Bellmare||time interval\\\\
    
            \textit{Input}: what did harry s truman work before he was president
                \newline \textit{Output}: when harry s truman president||time interval
            & \textit{Input}: Which album was released by the Smashing Pumpkins after Mike Byrne joined the band
                \newline \textit{Output}: When Mike Byrne joined Smashing Pumpkins||time interval\\
            \bottomrule
        \end{tabular}
    }
	\label{tab:prompts}
    \vspace*{-0.2cm}
\end{table*}

\section{Further implementation details}
\subsection{Training \faith on \tiq} 
\label{sec:training-on-tiq}
\faith requires train questions asking for temporal values
to answer intermediate questions.
Such questions exist in \timequestions,
but our new \benchmark benchmark only has implicit questions (by design).
We thus generate intermediate questions on the train and dev sets, using the implicit questions as input (similar as in Sec.~\ref{subsec:tqu}).
For these intermediate questions, the gold answer is the temporal value of the implicit part as annotated in the \benchmark benchmark,
resulting in <question, temporal value> pairs.
If the answer type of an intermediate question is a time interval, we create two questions asking for \phrase{start date} and \phrase{end date} respectively. 
We obtain $7{,}723$ such pairs from the \benchmark train set and $2{,}542$ questions from the dev set.

\subsection{TSF Construction}
\label{sec:distant-supervision}

\myparagraph{Training}
To obtain training data for the {TSF}s, we follow a similar distant supervision approach as in~\cite{christmann2023explainable}
for obtaining the target \textit{question entity} and \textit{question relation}.
We run the heterogeneous retriever (see Sec.~\ref{subsec:fer}) on the full input question,
which identifies entity mentions in the input, disambiguates these to KB-entities, and then 
retrieves information snippets for the KB-entities from heterogeneous sources.
If the retrieved information snippets for a KB-entity contain the annotated gold answer,
we annotate the corresponding entity mention as relevant question entity.
The remaining parts of the question are annotated as question relation.
The \textit{expected answer type} is
the most frequent (proxy for most prominent) KB-type of the gold answer.
The \textit{temporal signal} and the \textit{temporal category}
are looked up from the annotations in the benchmarks.
These individual parts of the \tsf are then combined and separated by pipes (``||''),
to obtain the target {\tsf}s that are used for training the TSF construction model (in Sec.~\ref{subsec:tqu}).
Details on the training configuration are provided in Sec.~\ref{sec:setup}.

\myparagraph{Inference}
The input to the BART model for TSF construction is the question. 
The output is the concatenation of the individual slots, separated by two pipes (``||''):
``\{entities\}||\{relation\}||\{expected answer type\}||\{temporal signal\}||\{temporal categorization\}''.
Example output for $q_1$=\phrase{Record company of Queen in 1975?}:
``Queen||Record company of in 1975||record company||overlap||non-implicit''.

\begin{table} \footnotesize
    \caption{{Performance of \temporalqa (with all sources) on questions from different source combinations in \benchmark ({test} set).}}
    \vspace*{-0.4cm}
    \newcolumntype{G}{>{\columncolor [gray] {0.90}}c}
    \newcolumntype{L}{>{}c}
    	\begin{tabular}{l G G G}

    		\toprule

                \textbf{Question sources combination} 
                & \textbf{P@1} & \textbf{MRR} & \textbf{Hit@5}
                 \\
    		    
            \midrule
        		\textbf{[Text; Infobox]} ($157$ questions)
           &	$0.573$  &	$0.644$   &	$0.752$
               \\

                \textbf{[KB; Text]} ($378$ questions)
                &	$0.519$  &	$0.634$   &	$0.770$
                  
                \\

                \textbf{[Infobox; KB]} ($43$ questions)
                &	$0.395$  &	$0.534$   &	$0.721$
                 \\

                \textbf{[Infobox; Text]} ($225$ questions)
                &	$0.476$  &	$0.595$   &	$0.760$
                \\
            
                \textbf{[KB; Infobox]} ($251$ questions)
                &	$0.478$  &	$0.598$   &	$0.753$
                 \\    

                \textbf{[Text; KB]} ($142$ questions)
                &	$0.359$  &	$0.509$   &	$0.725$
                \\ \midrule     
                 \textbf{[KB; KB]} ($127$ questions)
                 &	$0.582$  &	$0.676$   &	$0.807$
                 \\      
                 \textbf{[Text; Text]} ($306$ questions)
                  &	$0.598$  &	$0.679$   &	$0.787$
                 \\      
                \textbf{[Infobox; Infobox]} ($376$ questions)
                 &	$0.407$  &	$0.529$   &	$0.689$
                 \\ \midrule  
                 \textbf{All test questions} ($2{,}000$ questions)
                 &	$0.497$  &	$0.610$   &	$0.756$\\
    		\bottomrule
    	\end{tabular} 

    \label{tab:faith-input-sources}
    \vspace*{-0.2cm}
\end{table}

\subsection{Intermediate Question Generation}
\label{sec:seq2seq}

\myparagraph{Training}
To annotate {intermediate questions} (and its expected answer type) for implicit questions, we leverage in-context learning: we select and label $8$ questions from the train set, and give these pairs as context to the LLM (InstructGPT).
This way we annotate the remaining questions in the train and dev sets of \benchmark/\timequestions
resulting in $5{,}875$/$847$ instances in the train set and $1{,}949$/$287$ instances in the dev set.
On this data we fine-tune the BART model to be independent of GPT at runtime.
The prompts used for annotating the data can be found in Table~\ref{tab:prompts}. 
Training configuration is provided in Sec.~\ref{sec:setup}.

\myparagraph{Inference}
For generating the intermediate questions at runtime, we provide the implicit question as input to the trained BART model.
The output is the intermediate question that describes the implicit constraint,
and the expected answer type for this question, separated by two pipes:
``\{intermediate question\}||\{expected answer type\}''.
Example output for $q_3$=\phrase{Queen's record company when recording Bohemian Rhapsody?}:
``when Queen recording Bohemian Rhapsody||time interval''.


\subsection{Evidence Scoring} 
As the set of candidate information snippets after temporal pruning can still be large,
we use a re-ranker~\cite{nogueira2019passage} to prune out irrelevant candidates, based on cross-encodings obtained via
DistilRoBERTa\footnote{~\url{https://huggingface.co/distilroberta-base}.}.

\label{sec:scoring-model}
\myparagraph{Training}
The training data are the <question, information snippet> pairs,
annotated with either a positive label (in case the snippet contains a gold answer) or a negative label (otherwise).
We randomly sample $1$ positive <question, information snippet> pair from each knowledge source and $15$ negative pairs, for each question. 
We use the concatenation of {question entity},  the {question relation}, and the {expected answer type},
as present in the \tsf, to represent the question.
For fine-tuning the classifier on this data, we use AdamW as optimizer with a learning rate of $2\times$$10^{-5}$, batch size of $16$, weight decay of $0.01$, $4$ epochs, and a warm-up ratio of $0.1$.

\myparagraph{Inference}
We score each candidate information snippet (known to be temporally faithful)
and consider the top-$100$ information snippets as input for the final answering stage.

\section{Further experiments}

\subsection{Intrinsic Evaluation}
\label{sec:intrinsic}

\myparagraph{Temporal signal accuracy}
We measure the accuracy of the generated temporal signals (\textit{before}, \textit{after} or \textit{overlap}) in our \tsf construction.
On \timequestions the accuracy is {$93.0\%$}, and on \benchmark it is {$97.8\%$}.
The high accuracy scores indicate that our approach of generating the temporal signal is feasible.

\myparagraph{Temporal category accuracy}
We also measure the accuracy in predicting the temporal category, differentiating between implicit and non-implicit questions.
The accuracy is $98.9\%$ on \timequestions and $100\%$
on \tiq, which has only implicit questions.

\myparagraph{Performance of implicit question resolver}
The performance of the implicit question resolver is crucial for correctly answering implicit questions,
as the resulting temporal values are directly used
for pruning out evidence in the remainder of the pipeline.
As there might be several temporal values per question (due to multiple intermediate questions),
we measure macro-averaged precision, recall, and F1-score.
We conduct experiments on the test set of \benchmark, which has the ground-truth temporal values.

When using the top-$1$ answer candidate per intermediate question (default setting),
precision is $0.537$, recall is $0.562$ and F1-score is $0.525$.
When increasing the number of candidates to $3$,
precision is $0.294$, recall is $0.714$ and F1-score is $0.401$.
With the top-$5$ candidates,
precision is $0.196$, recall is $0.773$ and F1-score is $0.304$.

Recall improves as we consider more answers, since the resulting explicit temporal constraint is relaxed.
Hence, the evidence retained is noisier and may not satisfy the user-intended temporal constraints.
This can negatively affect the system's faithfulness.

\subsection{Additional Analysis}
\label{sec:add-analysis}

\begin{figure} [t]
     \includegraphics[width=0.7\columnwidth]
     {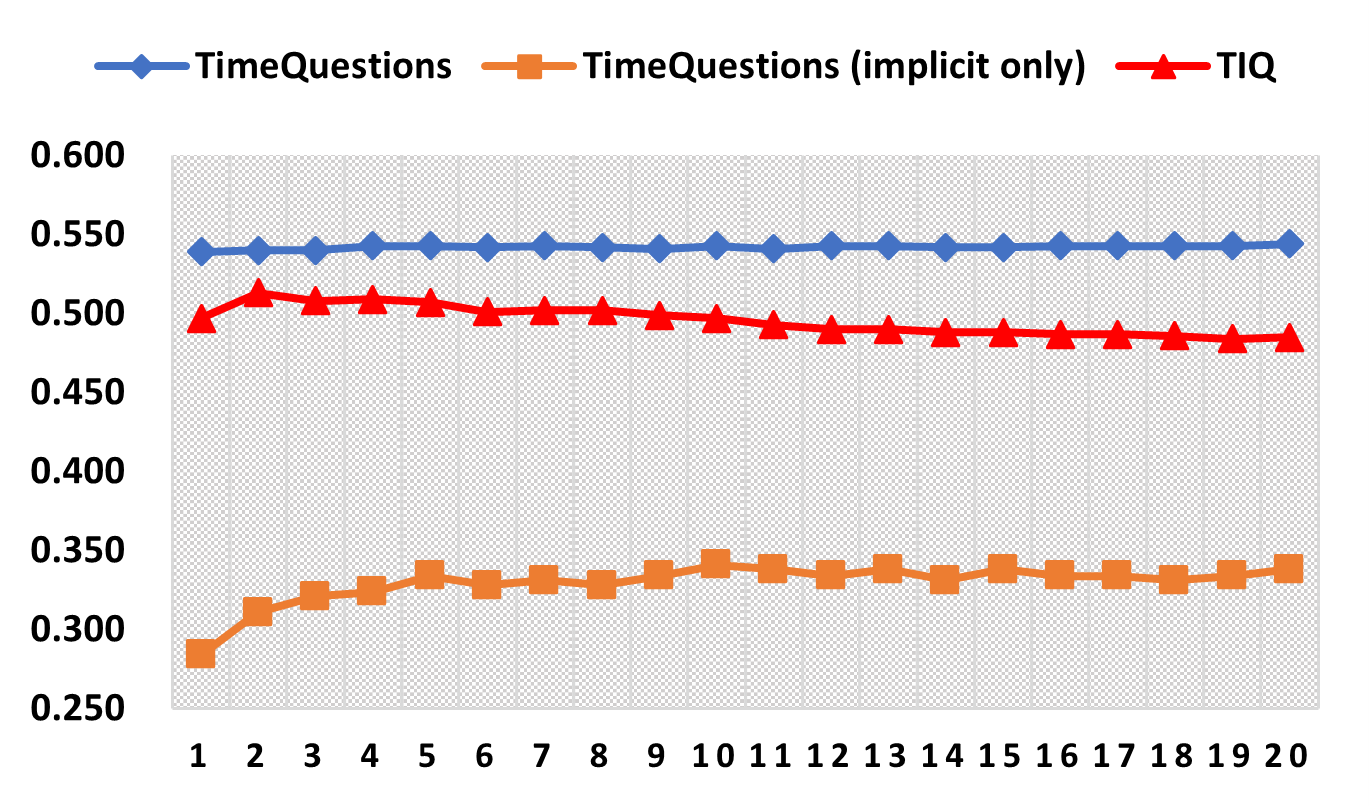}
     \vspace*{-0.5cm}
     \caption{P@1 of \faith when considering top-$k$ answers for the generated intermediate question(s) of implicit questions.}
     \label{fig:relax_tp}
     \vspace*{-0.1cm}
\end{figure}

\myparagraph{Relaxed temporal pruning}
We also compute the effect of the number of answer candidates in the implicit question resolver on the end-to-end performance (\textit{extrinsic} evaluation).
Fig.~\ref{fig:relax_tp} shows the results, varying the number of candidates $k$ from $1$ to $20$.
On \timequestions, we observe that P@1 improves gradually as $k$ increases until the set of candidate snippets converges resulting in a stable P@1.
On \benchmark, we only observe an improvement of the P@1 metric when increasing $k$ to $2$.
As $k$ increases further, more noisy candidate snippets are considered, resulting in a lower performance.

\myparagraph{Answer presence analysis}
We measure the answer presence after the initial heterogeneous retrieval,
and the effect of the subsequent pruning and scoring steps.
Answer presence is measured as the fraction of questions for which the gold answer
is present in the candidate set of information snippets~\cite{christmann2022beyond}.
These measurements are also used for the error analysis presented in Sec.~\ref{sec:analysis}.

We conduct the analysis on the test sets of \benchmark/\timequestions.
The answer presence after the heterogeneous retrieval is $0.984$/$0.861$.
After temporal pruning the answer presence drops to $0.872$/$0.741$.
Note that in this step, temporally inconsistent evidence is pruned out, enhancing the faithfulness of the approach.
In the evidence scoring stage (based on a cross-encoder),
the answer presence is mostly retained ($0.867$/$0.726$).
Inside the \explaignn pipeline, the answer presence
after evidence pruning is
$0.829$/$0.693$.

In general, as discussed in the error analysis, the key source of error is the fine-grained answer ranking step.

\myparagraph{\temporalqa performance on heterogeneous questions}
\tiq has questions originating from different source combinations (see Fig.~\ref{fig:benchmark-sources}).
Table~\ref{tab:faith-input-sources} shows how this affects the performance of \faith (all sources are used as input for answering).

Results demonstrate that \faith can deal with all of these questions, and there is no combination
for which \faith completely fails, indicating that \faith successfully incorporates all
heterogeneous sources during answering.
\temporalqa shows the best performance on questions from text and infoboxes (P@1 of $0.573$),
and the worst performance on questions from text and KB (P@1 of $0.359$).

%% file: 2024-www-fp-faith-cr.bbl

\begin{thebibliography}{65}


\ifx \showCODEN    \undefined \def \showCODEN     #1{\unskip}     \fi
\ifx \showDOI      \undefined \def \showDOI       #1{#1}\fi
\ifx \showISBNx    \undefined \def \showISBNx     #1{\unskip}     \fi
\ifx \showISBNxiii \undefined \def \showISBNxiii  #1{\unskip}     \fi
\ifx \showISSN     \undefined \def \showISSN      #1{\unskip}     \fi
\ifx \showLCCN     \undefined \def \showLCCN      #1{\unskip}     \fi
\ifx \shownote     \undefined \def \shownote      #1{#1}          \fi
\ifx \showarticletitle \undefined \def \showarticletitle #1{#1}   \fi
\ifx \showURL      \undefined \def \showURL       {\relax}        \fi
\providecommand\bibfield[2]{#2}
\providecommand\bibinfo[2]{#2}
\providecommand\natexlab[1]{#1}
\providecommand\showeprint[2][]{arXiv:#2}

\bibitem[Adlakha et~al\mbox{.}(2023)]%
        {adlakha2023evaluating}
\bibfield{author}{\bibinfo{person}{Vaibhav Adlakha}, \bibinfo{person}{Parishad BehnamGhader}, \bibinfo{person}{Xing~Han Lu}, \bibinfo{person}{Nicholas Meade}, {and} \bibinfo{person}{Siva Reddy}.} \bibinfo{year}{2023}\natexlab{}.
\newblock \showarticletitle{Evaluating Correctness and Faithfulness of Instruction-Following Models for Question Answering}. In \bibinfo{booktitle}{\emph{arXiv}}.
\newblock


\bibitem[Berant et~al\mbox{.}(2013)]%
        {berant2013semantic}
\bibfield{author}{\bibinfo{person}{Jonathan Berant}, \bibinfo{person}{Andrew Chou}, \bibinfo{person}{Roy Frostig}, {and} \bibinfo{person}{Percy Liang}.} \bibinfo{year}{2013}\natexlab{}.
\newblock \showarticletitle{Semantic Parsing on Freebase from Question-Answer Pairs}. In \bibinfo{booktitle}{\emph{EMNLP}}.
\newblock


\bibitem[Brown et~al\mbox{.}(2020)]%
        {brown2020language}
\bibfield{author}{\bibinfo{person}{Tom Brown}, \bibinfo{person}{Benjamin Mann}, \bibinfo{person}{Nick Ryder}, \bibinfo{person}{Melanie Subbiah}, \bibinfo{person}{Jared~D Kaplan}, \bibinfo{person}{Prafulla Dhariwal}, \bibinfo{person}{Arvind Neelakantan}, \bibinfo{person}{Pranav Shyam}, \bibinfo{person}{Girish Sastry}, \bibinfo{person}{Amanda Askell}, {et~al\mbox{.}}} \bibinfo{year}{2020}\natexlab{}.
\newblock \showarticletitle{Language Models are Few-Shot Learners}. In \bibinfo{booktitle}{\emph{NeurIPS}}.
\newblock


\bibitem[Cai et~al\mbox{.}(2022)]%
        {cai2022temporal}
\bibfield{author}{\bibinfo{person}{Borui Cai}, \bibinfo{person}{Yong Xiang}, \bibinfo{person}{Longxiang Gao}, \bibinfo{person}{He Zhang}, \bibinfo{person}{Yunfeng Li}, {and} \bibinfo{person}{Jianxin Li}.} \bibinfo{year}{2022}\natexlab{}.
\newblock \showarticletitle{Temporal Knowledge Graph Completion: A Survey}. In \bibinfo{booktitle}{\emph{arXiv}}.
\newblock


\bibitem[Campos et~al\mbox{.}(2014)]%
        {campos2014survey}
\bibfield{author}{\bibinfo{person}{Ricardo Campos}, \bibinfo{person}{Ga{\"e}l Dias}, \bibinfo{person}{Al{\'\i}pio~M Jorge}, {and} \bibinfo{person}{Adam Jatowt}.} \bibinfo{year}{2014}\natexlab{}.
\newblock \showarticletitle{Survey of Temporal Information Retrieval and Related Applications}.
\newblock \bibinfo{journal}{\emph{ACM Computing Surveys (CSUR)}} (\bibinfo{year}{2014}).
\newblock


\bibitem[Chang and Manning(2012)]%
        {chang2012sutime}
\bibfield{author}{\bibinfo{person}{Angel~X. Chang} {and} \bibinfo{person}{Christopher~D. Manning}.} \bibinfo{year}{2012}\natexlab{}.
\newblock \showarticletitle{{SUTime: A} Library for Recognizing and Normalizing Time Expressions}. In \bibinfo{booktitle}{\emph{LREC}}.
\newblock


\bibitem[Chen et~al\mbox{.}(2017)]%
        {chen2017reading}
\bibfield{author}{\bibinfo{person}{Danqi Chen}, \bibinfo{person}{Adam Fisch}, \bibinfo{person}{Jason Weston}, {and} \bibinfo{person}{Antoine Bordes}.} \bibinfo{year}{2017}\natexlab{}.
\newblock \showarticletitle{Reading Wikipedia to Answer Open-Domain Questions}. In \bibinfo{booktitle}{\emph{ACL}}.
\newblock


\bibitem[Chen et~al\mbox{.}(2021a)]%
        {chen2021open}
\bibfield{author}{\bibinfo{person}{Wenhu Chen}, \bibinfo{person}{Ming-Wei Chang}, \bibinfo{person}{Eva Schlinger}, \bibinfo{person}{William Wang}, {and} \bibinfo{person}{William~W Cohen}.} \bibinfo{year}{2021}\natexlab{a}.
\newblock \showarticletitle{Open Question Answering over Tables and Text}. In \bibinfo{booktitle}{\emph{ICLR}}.
\newblock


\bibitem[Chen et~al\mbox{.}(2021b)]%
        {chen2021dataset}
\bibfield{author}{\bibinfo{person}{Wenhu Chen}, \bibinfo{person}{Xinyi Wang}, {and} \bibinfo{person}{William~Yang Wang}.} \bibinfo{year}{2021}\natexlab{b}.
\newblock \showarticletitle{A Dataset for Answering Time-Sensitive Questions}. In \bibinfo{booktitle}{\emph{NeurIPS}}.
\newblock


\bibitem[Chen et~al\mbox{.}(2022)]%
        {chen2022temporal}
\bibfield{author}{\bibinfo{person}{Ziyang Chen}, \bibinfo{person}{Xiang Zhao}, \bibinfo{person}{Jinzhi Liao}, \bibinfo{person}{Xinyi Li}, {and} \bibinfo{person}{Evangelos Kanoulas}.} \bibinfo{year}{2022}\natexlab{}.
\newblock \showarticletitle{Temporal Knowledge Graph Question Qnswering via Subgraph Reasoning}.
\newblock \bibinfo{journal}{\emph{Knowledge-Based Systems}} (\bibinfo{year}{2022}).
\newblock


\bibitem[Christmann et~al\mbox{.}(2022a)]%
        {christmann2022beyond}
\bibfield{author}{\bibinfo{person}{Philipp Christmann}, \bibinfo{person}{Rishiraj Saha~Roy}, {and} \bibinfo{person}{Gerhard Weikum}.} \bibinfo{year}{2022}\natexlab{a}.
\newblock \showarticletitle{Beyond NED: Fast and Effective Search Space Reduction for Complex Question Answering over Knowledge Bases}. In \bibinfo{booktitle}{\emph{WSDM}}.
\newblock


\bibitem[Christmann et~al\mbox{.}(2022b)]%
        {christmann2022conversational}
\bibfield{author}{\bibinfo{person}{Philipp Christmann}, \bibinfo{person}{Rishiraj Saha~Roy}, {and} \bibinfo{person}{Gerhard Weikum}.} \bibinfo{year}{2022}\natexlab{b}.
\newblock \showarticletitle{Conversational Question Answering on Heterogeneous Sources}. In \bibinfo{booktitle}{\emph{SIGIR}}.
\newblock


\bibitem[Christmann et~al\mbox{.}(2023)]%
        {christmann2023explainable}
\bibfield{author}{\bibinfo{person}{Philipp Christmann}, \bibinfo{person}{Rishiraj Saha~Roy}, {and} \bibinfo{person}{Gerhard Weikum}.} \bibinfo{year}{2023}\natexlab{}.
\newblock \showarticletitle{Explainable Conversational Question Answering over Heterogeneous Sources via Iterative Graph Neural Networks}. In \bibinfo{booktitle}{\emph{SIGIR}}.
\newblock


\bibitem[Devlin et~al\mbox{.}(2019)]%
        {devlin2019bert}
\bibfield{author}{\bibinfo{person}{Jacob Devlin}, \bibinfo{person}{Ming-Wei Chang}, \bibinfo{person}{Kenton Lee}, {and} \bibinfo{person}{Kristina Toutanova}.} \bibinfo{year}{2019}\natexlab{}.
\newblock \showarticletitle{BERT: Pre-training of Deep Bidirectional Transformers for Language Understanding}. In \bibinfo{booktitle}{\emph{NAACL}}.
\newblock


\bibitem[Dhingra et~al\mbox{.}(2022)]%
        {dhingra2022time}
\bibfield{author}{\bibinfo{person}{Bhuwan Dhingra}, \bibinfo{person}{Jeremy~R Cole}, \bibinfo{person}{Julian~Martin Eisenschlos}, \bibinfo{person}{Daniel Gillick}, \bibinfo{person}{Jacob Eisenstein}, {and} \bibinfo{person}{William~W Cohen}.} \bibinfo{year}{2022}\natexlab{}.
\newblock \showarticletitle{Time-Aware Language Models as Temporal Knowledge Bases}. In \bibinfo{booktitle}{\emph{TACL}}.
\newblock


\bibitem[Ding et~al\mbox{.}(2022)]%
        {ding2022semantic}
\bibfield{author}{\bibinfo{person}{Wentao Ding}, \bibinfo{person}{Hao Chen}, \bibinfo{person}{Huayu Li}, {and} \bibinfo{person}{Yuzhong Qu}.} \bibinfo{year}{2022}\natexlab{}.
\newblock \showarticletitle{Semantic Framework based Query Generation for Temporal Question Answering over Knowledge Graphs}. In \bibinfo{booktitle}{\emph{EMNLP}}.
\newblock


\bibitem[Ferragina and Scaiella(2010)]%
        {ferragina2010tagme}
\bibfield{author}{\bibinfo{person}{Paolo Ferragina} {and} \bibinfo{person}{Ugo Scaiella}.} \bibinfo{year}{2010}\natexlab{}.
\newblock \showarticletitle{TAGME: On-the-fly Annotation of Short Text Fragments (by Wikipedia Entities)}. In \bibinfo{booktitle}{\emph{CIKM}}.
\newblock


\bibitem[Ferrucci(2012)]%
        {ferrucci2012introduction}
\bibfield{author}{\bibinfo{person}{David~A. Ferrucci}.} \bibinfo{year}{2012}\natexlab{}.
\newblock \showarticletitle{Introduction to "This is Watson"}.
\newblock \bibinfo{journal}{\emph{IBM Journal of Research and Development}} (\bibinfo{year}{2012}).
\newblock


\bibitem[Garcia-Duran et~al\mbox{.}(2018)]%
        {garcia2018learning}
\bibfield{author}{\bibinfo{person}{Alberto Garcia-Duran}, \bibinfo{person}{Sebastijan Duman{\v{c}}i{\'c}}, {and} \bibinfo{person}{Mathias Niepert}.} \bibinfo{year}{2018}\natexlab{}.
\newblock \showarticletitle{Learning Sequence Encoders for Temporal Knowledge Graph Completion}. In \bibinfo{booktitle}{\emph{EMNLP}}.
\newblock


\bibitem[Ho et~al\mbox{.}(2019)]%
        {ho2019qsearch}
\bibfield{author}{\bibinfo{person}{Vinh~Thinh Ho}, \bibinfo{person}{Yusra Ibrahim}, \bibinfo{person}{Koninika Pal}, \bibinfo{person}{Klaus Berberich}, {and} \bibinfo{person}{Gerhard Weikum}.} \bibinfo{year}{2019}\natexlab{}.
\newblock \showarticletitle{Qsearch: Answering Quantity Queries from Text}. In \bibinfo{booktitle}{\emph{ISWC}}.
\newblock


\bibitem[Izacard and Grave(2021)]%
        {izacard2021leveraging}
\bibfield{author}{\bibinfo{person}{Gautier Izacard} {and} \bibinfo{person}{{\'E}douard Grave}.} \bibinfo{year}{2021}\natexlab{}.
\newblock \showarticletitle{Leveraging Passage Retrieval with Generative Models for Open Domain Question Answering}. In \bibinfo{booktitle}{\emph{EACL}}.
\newblock


\bibitem[Jia et~al\mbox{.}(2018a)]%
        {jia2018tempquestions}
\bibfield{author}{\bibinfo{person}{Zhen Jia}, \bibinfo{person}{Abdalghani Abujabal}, \bibinfo{person}{Rishiraj Saha~Roy}, \bibinfo{person}{Jannik Str{\"o}tgen}, {and} \bibinfo{person}{Gerhard Weikum}.} \bibinfo{year}{2018}\natexlab{a}.
\newblock \showarticletitle{TempQuestions: A Benchmark for Temporal Question Answering}. In \bibinfo{booktitle}{\emph{HQA@WWW}}.
\newblock


\bibitem[Jia et~al\mbox{.}(2018b)]%
        {jia18tequila}
\bibfield{author}{\bibinfo{person}{Zhen Jia}, \bibinfo{person}{Abdalghani Abujabal}, \bibinfo{person}{Rishiraj Saha~Roy}, \bibinfo{person}{Jannik Str\"{o}tgen}, {and} \bibinfo{person}{Gerhard Weikum}.} \bibinfo{year}{2018}\natexlab{b}.
\newblock \showarticletitle{TEQUILA: Temporal Question Answering over Knowledge Bases}. In \bibinfo{booktitle}{\emph{CIKM}}.
\newblock


\bibitem[Jia et~al\mbox{.}(2021)]%
        {jia2021complex}
\bibfield{author}{\bibinfo{person}{Zhen Jia}, \bibinfo{person}{Soumajit Pramanik}, \bibinfo{person}{Rishiraj Saha~Roy}, {and} \bibinfo{person}{Gerhard Weikum}.} \bibinfo{year}{2021}\natexlab{}.
\newblock \showarticletitle{Complex Temporal Question Answering on Knowledge Graphs}. In \bibinfo{booktitle}{\emph{CIKM}}.
\newblock


\bibitem[Jiao et~al\mbox{.}(2022)]%
        {jiao2022improving}
\bibfield{author}{\bibinfo{person}{Songlin Jiao}, \bibinfo{person}{Zhenfang Zhu}, \bibinfo{person}{Wenqing Wu}, \bibinfo{person}{Zicheng Zuo}, \bibinfo{person}{Jiangtao Qi}, \bibinfo{person}{Wenling Wang}, \bibinfo{person}{Guangyuan Zhang}, {and} \bibinfo{person}{Peiyu Liu}.} \bibinfo{year}{2022}\natexlab{}.
\newblock \showarticletitle{An Improving Reasoning Network for Complex Question Answering over Temporal Knowledge Graphs}. In \bibinfo{booktitle}{\emph{Applied Intelligence}}. \bibinfo{publisher}{Springer}.
\newblock


\bibitem[Leblay and Chekol(2018)]%
        {leblay2018deriving}
\bibfield{author}{\bibinfo{person}{Julien Leblay} {and} \bibinfo{person}{Melisachew~Wudage Chekol}.} \bibinfo{year}{2018}\natexlab{}.
\newblock \showarticletitle{Deriving Validity Time in Knowledge Graph}. In \bibinfo{booktitle}{\emph{TempWeb@WWW}}.
\newblock


\bibitem[Lewis et~al\mbox{.}(2020)]%
        {lewis2020bart}
\bibfield{author}{\bibinfo{person}{Mike Lewis}, \bibinfo{person}{Yinhan Liu}, \bibinfo{person}{Naman Goyal}, \bibinfo{person}{Marjan Ghazvininejad}, \bibinfo{person}{Abdelrahman Mohamed}, \bibinfo{person}{Omer Levy}, \bibinfo{person}{Veselin Stoyanov}, {and} \bibinfo{person}{Luke Zettlemoyer}.} \bibinfo{year}{2020}\natexlab{}.
\newblock \showarticletitle{BART: Denoising Sequence-to-Sequence Pre-training for Natural Language Generation, Translation, and Comprehension}. In \bibinfo{booktitle}{\emph{ACL}}.
\newblock


\bibitem[Long et~al\mbox{.}(2022)]%
        {long2022complex}
\bibfield{author}{\bibinfo{person}{Shaonan Long}, \bibinfo{person}{Jinzhi Liao}, \bibinfo{person}{Shiyu Yang}, \bibinfo{person}{Xiang Zhao}, {and} \bibinfo{person}{Xuemin Lin}.} \bibinfo{year}{2022}\natexlab{}.
\newblock \showarticletitle{Complex Question Answering Over Temporal Knowledge Graphs}. In \bibinfo{booktitle}{\emph{WISE}}.
\newblock


\bibitem[Ma et~al\mbox{.}(2022)]%
        {ma2021open}
\bibfield{author}{\bibinfo{person}{Kaixin Ma}, \bibinfo{person}{Hao Cheng}, \bibinfo{person}{Xiaodong Liu}, \bibinfo{person}{Eric Nyberg}, {and} \bibinfo{person}{Jianfeng Gao}.} \bibinfo{year}{2022}\natexlab{}.
\newblock \showarticletitle{Open Domain Question Answering with A Unified Knowledge Interface}. In \bibinfo{booktitle}{\emph{ACL}}.
\newblock


\bibitem[Mallen et~al\mbox{.}(2023)]%
        {mallen2023not}
\bibfield{author}{\bibinfo{person}{Alex Mallen}, \bibinfo{person}{Akari Asai}, \bibinfo{person}{Victor Zhong}, \bibinfo{person}{Rajarshi Das}, \bibinfo{person}{Daniel Khashabi}, {and} \bibinfo{person}{Hannaneh Hajishirzi}.} \bibinfo{year}{2023}\natexlab{}.
\newblock \showarticletitle{When not to Trust Language Models: Investigating Effectiveness of Parametric and Non-Parametric Memories}. In \bibinfo{booktitle}{\emph{ACL}}.
\newblock


\bibitem[Mavromatis et~al\mbox{.}(2022)]%
        {mavromatis2022tempoqr}
\bibfield{author}{\bibinfo{person}{Costas Mavromatis}, \bibinfo{person}{Prasanna~Lakkur Subramanyam}, \bibinfo{person}{Vassilis~N Ioannidis}, \bibinfo{person}{Adesoji Adeshina}, \bibinfo{person}{Phillip~R Howard}, \bibinfo{person}{Tetiana Grinberg}, \bibinfo{person}{Nagib Hakim}, {and} \bibinfo{person}{George Karypis}.} \bibinfo{year}{2022}\natexlab{}.
\newblock \showarticletitle{TempoQR: Temporal Question Reasoning over Knowledge Graphs}. In \bibinfo{booktitle}{\emph{AAAI}}.
\newblock


\bibitem[Messner et~al\mbox{.}(2022)]%
        {messner2022temporal}
\bibfield{author}{\bibinfo{person}{Johannes Messner}, \bibinfo{person}{Ralph Abboud}, {and} \bibinfo{person}{Ismail~Ilkan Ceylan}.} \bibinfo{year}{2022}\natexlab{}.
\newblock \showarticletitle{Temporal Knowledge Graph Completion Using Box Embeddings}. In \bibinfo{booktitle}{\emph{AAAI}}.
\newblock


\bibitem[Min et~al\mbox{.}(2023)]%
        {min2023factscore}
\bibfield{author}{\bibinfo{person}{Sewon Min}, \bibinfo{person}{Kalpesh Krishna}, \bibinfo{person}{Xinxi Lyu}, \bibinfo{person}{Mike Lewis}, \bibinfo{person}{Wen-tau Yih}, \bibinfo{person}{Pang~Wei Koh}, \bibinfo{person}{Mohit Iyyer}, \bibinfo{person}{Luke Zettlemoyer}, {and} \bibinfo{person}{Hannaneh Hajishirzi}.} \bibinfo{year}{2023}\natexlab{}.
\newblock \showarticletitle{FActScore: Fine-grained Atomic Evaluation of Factual Precision in Long Form Text Generation}. In \bibinfo{booktitle}{\emph{EMNLP}}.
\newblock


\bibitem[Neelam et~al\mbox{.}(2022)]%
        {neelam-etal-2022-sygma}
\bibfield{author}{\bibinfo{person}{Sumit Neelam}, \bibinfo{person}{Udit Sharma}, \bibinfo{person}{Hima Karanam}, \bibinfo{person}{Shajith Ikbal}, \bibinfo{person}{Pavan Kapanipathi}, \bibinfo{person}{Ibrahim Abdelaziz}, \bibinfo{person}{Nandana Mihindukulasooriya}, \bibinfo{person}{Young-Suk Lee}, \bibinfo{person}{Santosh Srivastava}, \bibinfo{person}{Cezar Pendus}, {et~al\mbox{.}}} \bibinfo{year}{2022}\natexlab{}.
\newblock \showarticletitle{SYGMA: System for Generalizable Modular Question Answering over Knowledge Bases}. In \bibinfo{booktitle}{\emph{Findings of the Association for Computational Linguistics: EMNLP}}.
\newblock


\bibitem[Ning et~al\mbox{.}(2020)]%
        {ning2020torque}
\bibfield{author}{\bibinfo{person}{Qiang Ning}, \bibinfo{person}{Hao Wu}, \bibinfo{person}{Rujun Han}, \bibinfo{person}{Nanyun Peng}, \bibinfo{person}{Matt Gardner}, {and} \bibinfo{person}{Dan Roth}.} \bibinfo{year}{2020}\natexlab{}.
\newblock \showarticletitle{TORQUE: A Reading Comprehension Dataset of Temporal Ordering Questions}. In \bibinfo{booktitle}{\emph{EMNLP}}.
\newblock


\bibitem[Nogueira and Cho(2019)]%
        {nogueira2019passage}
\bibfield{author}{\bibinfo{person}{Rodrigo Nogueira} {and} \bibinfo{person}{Kyunghyun Cho}.} \bibinfo{year}{2019}\natexlab{}.
\newblock \showarticletitle{Passage Re-ranking with BERT}. In \bibinfo{booktitle}{\emph{arXiv}}.
\newblock


\bibitem[OpenAI(2023)]%
        {openai2023gpt4}
\bibfield{author}{\bibinfo{person}{OpenAI}.} \bibinfo{year}{2023}\natexlab{}.
\newblock \showarticletitle{GPT-4 Technical Report}. In \bibinfo{booktitle}{\emph{arXiv}}.
\newblock


\bibitem[O\u{g}uz et~al\mbox{.}(2022)]%
        {oguz2021unikqa}
\bibfield{author}{\bibinfo{person}{Barlas O\u{g}uz}, \bibinfo{person}{Xilun Chen}, \bibinfo{person}{Vladimir Karpukhin}, \bibinfo{person}{Stan Peshterliev}, \bibinfo{person}{Dmytro Okhonko}, \bibinfo{person}{Michael Schlichtkrull}, \bibinfo{person}{Sonal Gupta}, \bibinfo{person}{Yashar Mehdad}, {and} \bibinfo{person}{Scott Yih}.} \bibinfo{year}{2022}\natexlab{}.
\newblock \showarticletitle{UniK-QA: Unified Representations of Structured and Unstructured Knowledge for Open-Domain Question Answering}. In \bibinfo{booktitle}{\emph{NAACL-HLT}}.
\newblock


\bibitem[Ouyang et~al\mbox{.}(2022)]%
        {ouyang2022training}
\bibfield{author}{\bibinfo{person}{Long Ouyang}, \bibinfo{person}{Jeffrey Wu}, \bibinfo{person}{Xu Jiang}, \bibinfo{person}{Diogo Almeida}, \bibinfo{person}{Carroll Wainwright}, \bibinfo{person}{Pamela Mishkin}, \bibinfo{person}{Chong Zhang}, \bibinfo{person}{Sandhini Agarwal}, \bibinfo{person}{Katarina Slama}, \bibinfo{person}{Alex Ray}, {et~al\mbox{.}}} \bibinfo{year}{2022}\natexlab{}.
\newblock \showarticletitle{Training Language Models to Follow Instructions with Human Feedback}. In \bibinfo{booktitle}{\emph{NeurIPS}}.
\newblock


\bibitem[Park et~al\mbox{.}(2022)]%
        {park2022evokg}
\bibfield{author}{\bibinfo{person}{Namyong Park}, \bibinfo{person}{Fuchen Liu}, \bibinfo{person}{Purvanshi Mehta}, \bibinfo{person}{Dana Cristofor}, \bibinfo{person}{Christos Faloutsos}, {and} \bibinfo{person}{Yuxiao Dong}.} \bibinfo{year}{2022}\natexlab{}.
\newblock \showarticletitle{Evokg: Jointly Modeling Event Time and Network Structure for Reasoning over Temporal Knowledge Graphs}. In \bibinfo{booktitle}{\emph{WSDM}}.
\newblock


\bibitem[Petroni et~al\mbox{.}(2019)]%
        {petroni2019language}
\bibfield{author}{\bibinfo{person}{Fabio Petroni}, \bibinfo{person}{Tim Rockt{\"a}schel}, \bibinfo{person}{Sebastian Riedel}, \bibinfo{person}{Patrick Lewis}, \bibinfo{person}{Anton Bakhtin}, \bibinfo{person}{Yuxiang Wu}, {and} \bibinfo{person}{Alexander Miller}.} \bibinfo{year}{2019}\natexlab{}.
\newblock \showarticletitle{Language Models as Knowledge Bases?}. In \bibinfo{booktitle}{\emph{EMNLP-IJCNLP}}.
\newblock


\bibitem[Pramanik et~al\mbox{.}(2021)]%
        {pramanik2021uniqorn}
\bibfield{author}{\bibinfo{person}{Soumajit Pramanik}, \bibinfo{person}{Jesujoba Alabi}, \bibinfo{person}{Rishiraj~Saha Roy}, {and} \bibinfo{person}{Gerhard Weikum}.} \bibinfo{year}{2021}\natexlab{}.
\newblock \showarticletitle{{UNIQORN:} Unified Question Answering over {RDF} Knowledge Graphs and Natural Language Text}. In \bibinfo{booktitle}{\emph{arXiv}}.
\newblock


\bibitem[Raffel et~al\mbox{.}(2020)]%
        {raffel2020exploring}
\bibfield{author}{\bibinfo{person}{Colin Raffel}, \bibinfo{person}{Noam Shazeer}, \bibinfo{person}{Adam Roberts}, \bibinfo{person}{Katherine Lee}, \bibinfo{person}{Sharan Narang}, \bibinfo{person}{Michael Matena}, \bibinfo{person}{Yanqi Zhou}, \bibinfo{person}{Wei Li}, {and} \bibinfo{person}{Peter~J Liu}.} \bibinfo{year}{2020}\natexlab{}.
\newblock \showarticletitle{Exploring the Limits of Transfer Learning with a Unified Text-to-Text Transformer}. In \bibinfo{booktitle}{\emph{JMLR}}.
\newblock


\bibitem[Rajpurkar et~al\mbox{.}(2016)]%
        {rajpurkar2016squad}
\bibfield{author}{\bibinfo{person}{Pranav Rajpurkar}, \bibinfo{person}{Jian Zhang}, \bibinfo{person}{Konstantin Lopyrev}, {and} \bibinfo{person}{Percy Liang}.} \bibinfo{year}{2016}\natexlab{}.
\newblock \showarticletitle{SQuAD: 100,000+ Questions for Machine Comprehension of Text}. In \bibinfo{booktitle}{\emph{EMNLP}}.
\newblock


\bibitem[Reimers and Gurevych(2019)]%
        {reimers-2019-sentence-bert}
\bibfield{author}{\bibinfo{person}{Nils Reimers} {and} \bibinfo{person}{Iryna Gurevych}.} \bibinfo{year}{2019}\natexlab{}.
\newblock \showarticletitle{Sentence-BERT: Sentence Embeddings using Siamese BERT-Networks}. In \bibinfo{booktitle}{\emph{EMNLP}}.
\newblock


\bibitem[Roy and Anand(2022)]%
        {saharoy2022question}
\bibfield{author}{\bibinfo{person}{Rishiraj~Saha Roy} {and} \bibinfo{person}{Avishek Anand}.} \bibinfo{year}{2022}\natexlab{}.
\newblock \bibinfo{booktitle}{\emph{Question Answering for the Curated Web: Tasks and Methods in QA over Knowledge Bases and Text Collections}}.
\newblock \bibinfo{publisher}{Springer}.
\newblock


\bibitem[Savenkov and Agichtein(2016)]%
        {savenkov2016knowledge}
\bibfield{author}{\bibinfo{person}{Denis Savenkov} {and} \bibinfo{person}{Eugene Agichtein}.} \bibinfo{year}{2016}\natexlab{}.
\newblock \showarticletitle{When a Knowledge Base Is Not Enough: Question Answering over Knowledge Bases with External Text Data}. In \bibinfo{booktitle}{\emph{SIGIR}}.
\newblock


\bibitem[Saxena et~al\mbox{.}(2021)]%
        {saxena2021question}
\bibfield{author}{\bibinfo{person}{Apoorv Saxena}, \bibinfo{person}{Soumen Chakrabarti}, {and} \bibinfo{person}{Partha Talukdar}.} \bibinfo{year}{2021}\natexlab{}.
\newblock \showarticletitle{Question Answering Over Temporal Knowledge Graphs}. In \bibinfo{booktitle}{\emph{ACL}}.
\newblock


\bibitem[Shang et~al\mbox{.}(2022)]%
        {shang2022improving}
\bibfield{author}{\bibinfo{person}{Chao Shang}, \bibinfo{person}{Guangtao Wang}, \bibinfo{person}{Peng Qi}, {and} \bibinfo{person}{Jing Huang}.} \bibinfo{year}{2022}\natexlab{}.
\newblock \showarticletitle{Improving Time Sensitivity for Question Answering over Temporal Knowledge Graphs}. In \bibinfo{booktitle}{\emph{ACL}}.
\newblock


\bibitem[Sharma et~al\mbox{.}(2023)]%
        {sharma2022twirgcn}
\bibfield{author}{\bibinfo{person}{Aditya Sharma}, \bibinfo{person}{Apoorv Saxena}, \bibinfo{person}{Chitrank Gupta}, \bibinfo{person}{Seyed~Mehran Kazemi}, \bibinfo{person}{Partha Talukdar}, {and} \bibinfo{person}{Soumen Chakrabarti}.} \bibinfo{year}{2023}\natexlab{}.
\newblock \showarticletitle{TwiRGCN: Temporally Weighted Graph Convolution for Question Answering over Temporal Knowledge Graphs}. In \bibinfo{booktitle}{\emph{EACL}}.
\newblock


\bibitem[Sun et~al\mbox{.}(2019)]%
        {sun2019pullnet}
\bibfield{author}{\bibinfo{person}{Haitian Sun}, \bibinfo{person}{Tania Bedrax-Weiss}, {and} \bibinfo{person}{William Cohen}.} \bibinfo{year}{2019}\natexlab{}.
\newblock \showarticletitle{PullNet: Open Domain Question Answering with Iterative Retrieval on Knowledge Bases and Text}. In \bibinfo{booktitle}{\emph{EMNLP-IJCNLP}}.
\newblock


\bibitem[Sun et~al\mbox{.}(2018)]%
        {sun2018open}
\bibfield{author}{\bibinfo{person}{Haitian Sun}, \bibinfo{person}{Bhuwan Dhingra}, \bibinfo{person}{Manzil Zaheer}, \bibinfo{person}{Kathryn Mazaitis}, \bibinfo{person}{Ruslan Salakhutdinov}, {and} \bibinfo{person}{William~W. Cohen}.} \bibinfo{year}{2018}\natexlab{}.
\newblock \showarticletitle{Open Domain Question Answering Using Early Fusion of Knowledge Bases and Text}. In \bibinfo{booktitle}{\emph{EMNLP}}.
\newblock


\bibitem[Tan et~al\mbox{.}(2023)]%
        {tan2023towards}
\bibfield{author}{\bibinfo{person}{Qingyu Tan}, \bibinfo{person}{Hwee~Tou Ng}, {and} \bibinfo{person}{Lidong Bing}.} \bibinfo{year}{2023}\natexlab{}.
\newblock \showarticletitle{Towards Benchmarking and Improving the Temporal Reasoning Capability of Large Language Models}. In \bibinfo{booktitle}{\emph{ACL}}.
\newblock


\bibitem[Veselovsky et~al\mbox{.}(2023)]%
        {veselovsky2023artificial}
\bibfield{author}{\bibinfo{person}{Veniamin Veselovsky}, \bibinfo{person}{Manoel~Horta Ribeiro}, {and} \bibinfo{person}{Robert West}.} \bibinfo{year}{2023}\natexlab{}.
\newblock \showarticletitle{Artificial Artificial Artificial Intelligence: Crowd Workers Widely Use Large Language Models for Text Production Tasks}. In \bibinfo{booktitle}{\emph{arXiv}}.
\newblock


\bibitem[Vrande{\v{c}}i{\'c} and Kr{\"o}tzsch(2014)]%
        {vrandevcic2014wikidata}
\bibfield{author}{\bibinfo{person}{Denny Vrande{\v{c}}i{\'c}} {and} \bibinfo{person}{Markus Kr{\"o}tzsch}.} \bibinfo{year}{2014}\natexlab{}.
\newblock \showarticletitle{Wikidata: A Free Collaborative Knowledgebase}.
\newblock \bibinfo{journal}{\emph{CACM}} (\bibinfo{year}{2014}).
\newblock


\bibitem[Wu et~al\mbox{.}(2020a)]%
        {wu2020temp}
\bibfield{author}{\bibinfo{person}{Jiapeng Wu}, \bibinfo{person}{Meng Cao}, \bibinfo{person}{Jackie Chi~Kit Cheung}, {and} \bibinfo{person}{William~L Hamilton}.} \bibinfo{year}{2020}\natexlab{a}.
\newblock \showarticletitle{TeMP: Temporal Message Passing for Temporal Knowledge Graph Completion}. In \bibinfo{booktitle}{\emph{EMNLP}}.
\newblock


\bibitem[Wu et~al\mbox{.}(2020b)]%
        {wu2020introducing}
\bibfield{author}{\bibinfo{person}{Wenqing Wu}, \bibinfo{person}{Zhenfang Zhu}, \bibinfo{person}{Qiang Lu}, \bibinfo{person}{Dianyuan Zhang}, {and} \bibinfo{person}{Qiangqiang Guo}.} \bibinfo{year}{2020}\natexlab{b}.
\newblock \showarticletitle{Introducing External Knowledge to Answer Questions with Implicit Temporal Constraints over Knowledge Base}.
\newblock \bibinfo{journal}{\emph{Future Internet}} (\bibinfo{year}{2020}).
\newblock


\bibitem[Xiao et~al\mbox{.}(2022)]%
        {xiao2022modeling}
\bibfield{author}{\bibinfo{person}{Yao Xiao}, \bibinfo{person}{Guangyou Zhou}, {and} \bibinfo{person}{Jin Liu}.} \bibinfo{year}{2022}\natexlab{}.
\newblock \showarticletitle{Modeling Temporal-Sensitive Information for Complex Question Answering over Knowledge Graphs}. In \bibinfo{booktitle}{\emph{NLPCC}}.
\newblock


\bibitem[Xu et~al\mbox{.}(2021)]%
        {xu2021temporal}
\bibfield{author}{\bibinfo{person}{Chengjin Xu}, \bibinfo{person}{Yung-Yu Chen}, \bibinfo{person}{Mojtaba Nayyeri}, {and} \bibinfo{person}{Jens Lehmann}.} \bibinfo{year}{2021}\natexlab{}.
\newblock \showarticletitle{Temporal Knowledge Graph Completion using a Linear Temporal Regularizer and Multivector Embeddings}. In \bibinfo{booktitle}{\emph{NAACL}}.
\newblock


\bibitem[Xu et~al\mbox{.}(2016a)]%
        {xu2016hybrid}
\bibfield{author}{\bibinfo{person}{Kun Xu}, \bibinfo{person}{Yansong Feng}, \bibinfo{person}{Songfang Huang}, {and} \bibinfo{person}{Dongyan Zhao}.} \bibinfo{year}{2016}\natexlab{a}.
\newblock \showarticletitle{Hybrid Question Answering over Knowledge Base and Free Text}. In \bibinfo{booktitle}{\emph{COLING}}.
\newblock


\bibitem[Xu et~al\mbox{.}(2016b)]%
        {xu2016question}
\bibfield{author}{\bibinfo{person}{Kun Xu}, \bibinfo{person}{Siva Reddy}, \bibinfo{person}{Yansong Feng}, \bibinfo{person}{Songfang Huang}, {and} \bibinfo{person}{Dongyan Zhao}.} \bibinfo{year}{2016}\natexlab{b}.
\newblock \showarticletitle{Question Answering on Freebase via Relation Extraction and Textual Evidence}. In \bibinfo{booktitle}{\emph{ACL}}.
\newblock


\bibitem[Yahya et~al\mbox{.}(2012)]%
        {yahya2012natural}
\bibfield{author}{\bibinfo{person}{Mohamed Yahya}, \bibinfo{person}{Klaus Berberich}, \bibinfo{person}{Shady Elbassuoni}, \bibinfo{person}{Maya Ramanath}, \bibinfo{person}{Volker Tresp}, {and} \bibinfo{person}{Gerhard Weikum}.} \bibinfo{year}{2012}\natexlab{}.
\newblock \showarticletitle{Natural Language Questions for the Web of Data}. In \bibinfo{booktitle}{\emph{EMNLP}}.
\newblock


\bibitem[Yao et~al\mbox{.}(2022)]%
        {yao2022terqa}
\bibfield{author}{\bibinfo{person}{Junping Yao}, \bibinfo{person}{Yijing Wang}, \bibinfo{person}{Xiaojun Li}, \bibinfo{person}{Cong Yuan}, {and} \bibinfo{person}{Kaiyuan Cheng}.} \bibinfo{year}{2022}\natexlab{}.
\newblock \showarticletitle{TERQA: Question Answering over Knowledge Graph Considering Precise Dependencies of Temporal Information on Vectors}.
\newblock \bibinfo{journal}{\emph{Displays}} (\bibinfo{year}{2022}).
\newblock


\bibitem[Yih et~al\mbox{.}(2015)]%
        {yih2015semantic}
\bibfield{author}{\bibinfo{person}{Scott Wen-tau Yih}, \bibinfo{person}{Ming-Wei Chang}, \bibinfo{person}{Xiaodong He}, {and} \bibinfo{person}{Jianfeng Gao}.} \bibinfo{year}{2015}\natexlab{}.
\newblock \showarticletitle{Semantic Parsing via Staged Query Graph Generation: Question Answering with Knowledge Base}. In \bibinfo{booktitle}{\emph{ACL-IJCNLP}}.
\newblock


\bibitem[Zhang et~al\mbox{.}(2010)]%
        {zhang2010learning}
\bibfield{author}{\bibinfo{person}{Ruiqiang Zhang}, \bibinfo{person}{Yuki Konda}, \bibinfo{person}{Anlei Dong}, \bibinfo{person}{Pranam Kolari}, \bibinfo{person}{Yi Chang}, {and} \bibinfo{person}{Zhaohui Zheng}.} \bibinfo{year}{2010}\natexlab{}.
\newblock \showarticletitle{Learning Recurrent Event Queries for Web Search}. In \bibinfo{booktitle}{\emph{EMNLP}}.
\newblock


\end{thebibliography}
